\documentclass[conference]{IEEEtran}
\usepackage{cite}
\usepackage{graphicx}
\usepackage{psfrag}
\usepackage{url}
\usepackage{amsmath}
\usepackage{array}
\usepackage{amssymb}
\usepackage{amsfonts}
\usepackage{graphicx}
\usepackage{algorithm}
\usepackage{algpseudocode}
\newtheorem{proposition}{Proposition}
\newtheorem{lemma}{Lemma}

\newtheorem{definition}{Definition}
\newtheorem{theorem}{Theorem}

\newtheorem{example}{Example}

\usepackage{setspace}
\usepackage{amscd}
\usepackage{mathrsfs}
\usepackage{epsfig}
\usepackage{color}
\usepackage{textcomp}
\usepackage{multirow}
\usepackage{caption}
\usepackage{subcaption}
\usepackage{threeparttable}

\title{Physical Layer Network Coding for the $K$-user Multiple Access Relay Channel}
\begin{document}

\author{
\authorblockN{Vijayvaradharaj T. Muralidharan and B. Sundar Rajan}
\authorblockA{Dept. of ECE, IISc, Bangalore 560012, India, Email:{$\lbrace$tmvijay, bsrajan$\rbrace$}@ece.iisc.ernet.in
}
}

\maketitle
\begin{abstract}

A Physical layer Network Coding (PNC) scheme is proposed for the $K$-user wireless Multiple Access Relay Channel (MARC), in which $K$ source nodes transmit their messages to the destination node $D$ with the help of a relay node $R.$ The proposed PNC scheme involves two transmission phases: (i) Phase 1 during which the source nodes transmit, the relay node and the destination node receive and (ii) Phase 2 during which the source nodes and the relay node transmit, and the destination node receives. At the end of Phase 1, the relay node decodes the messages of the source nodes and during Phase 2 transmits a many-to-one function of the decoded messages. Wireless networks in which the relay node decodes, suffer from loss of diversity order if the decoder at the destination is not chosen properly. A novel decoder is proposed for the PNC scheme, which offers the maximum possible diversity order of $2,$ for a proper choice of certain parameters and the network coding map. Specifically, the network coding map used at the relay is chosen to be a $K$-dimensional Latin Hypercube, in order to ensure the maximum diversity order of $2.$ Also, it is shown that the proposed decoder can be implemented by a fast decoding algorithm. Simulation results presented for the 3-user and 4-user MARC show that the proposed scheme offers a large gain over the existing scheme for the $K$-user MARC.
\end{abstract}

\section{Background and Preliminaries}
\begin{figure}[htbp]
\centering
\begin{subfigure}[]{1.7in}
\includegraphics[totalheight=1.45in,width=1.75in]{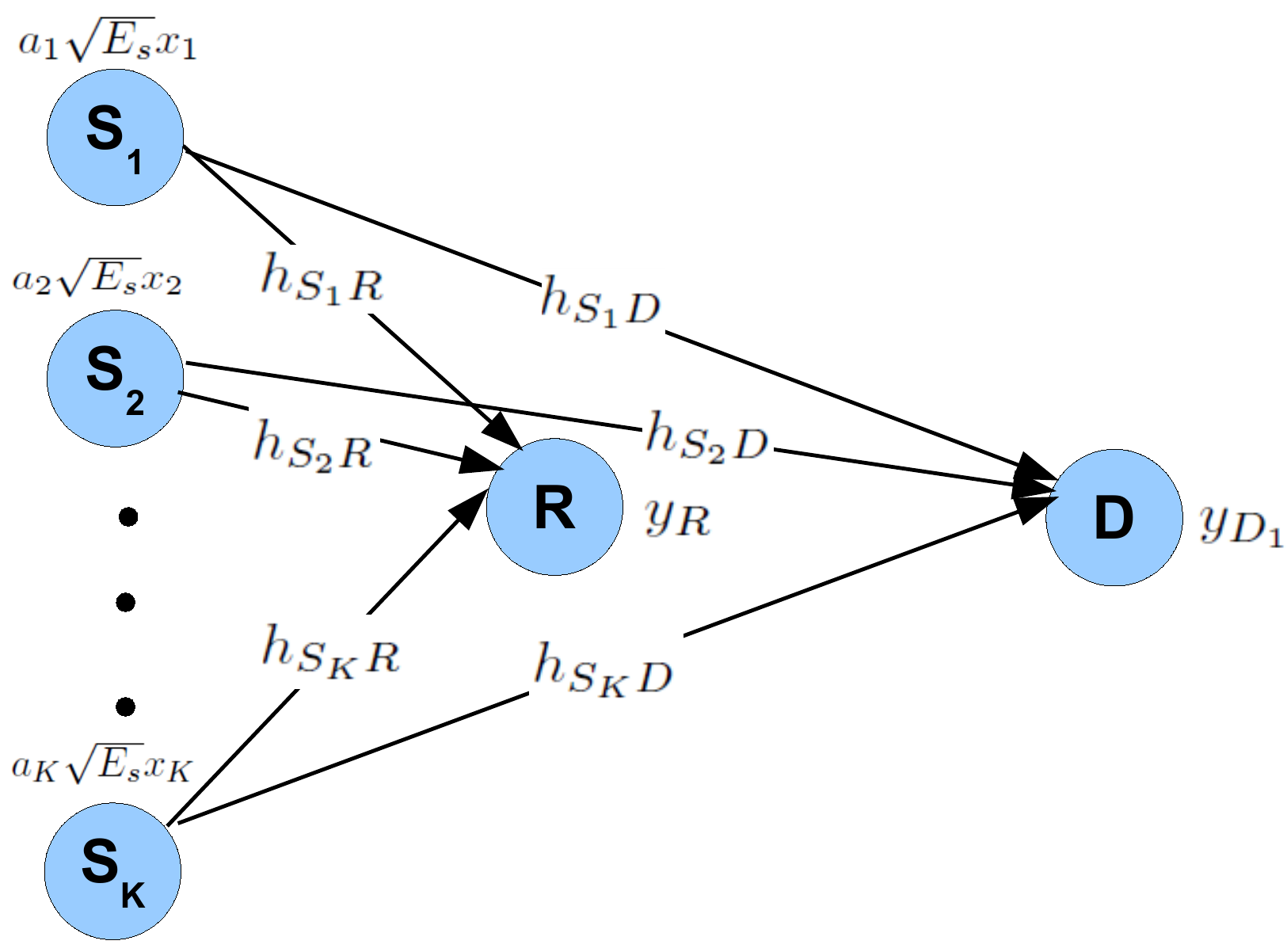}
\label{fig:phase1}
\caption{Phase 1}
\end{subfigure}
\begin{subfigure}[]{1.7in}
\includegraphics[totalheight=1.45in,width=1.75in]{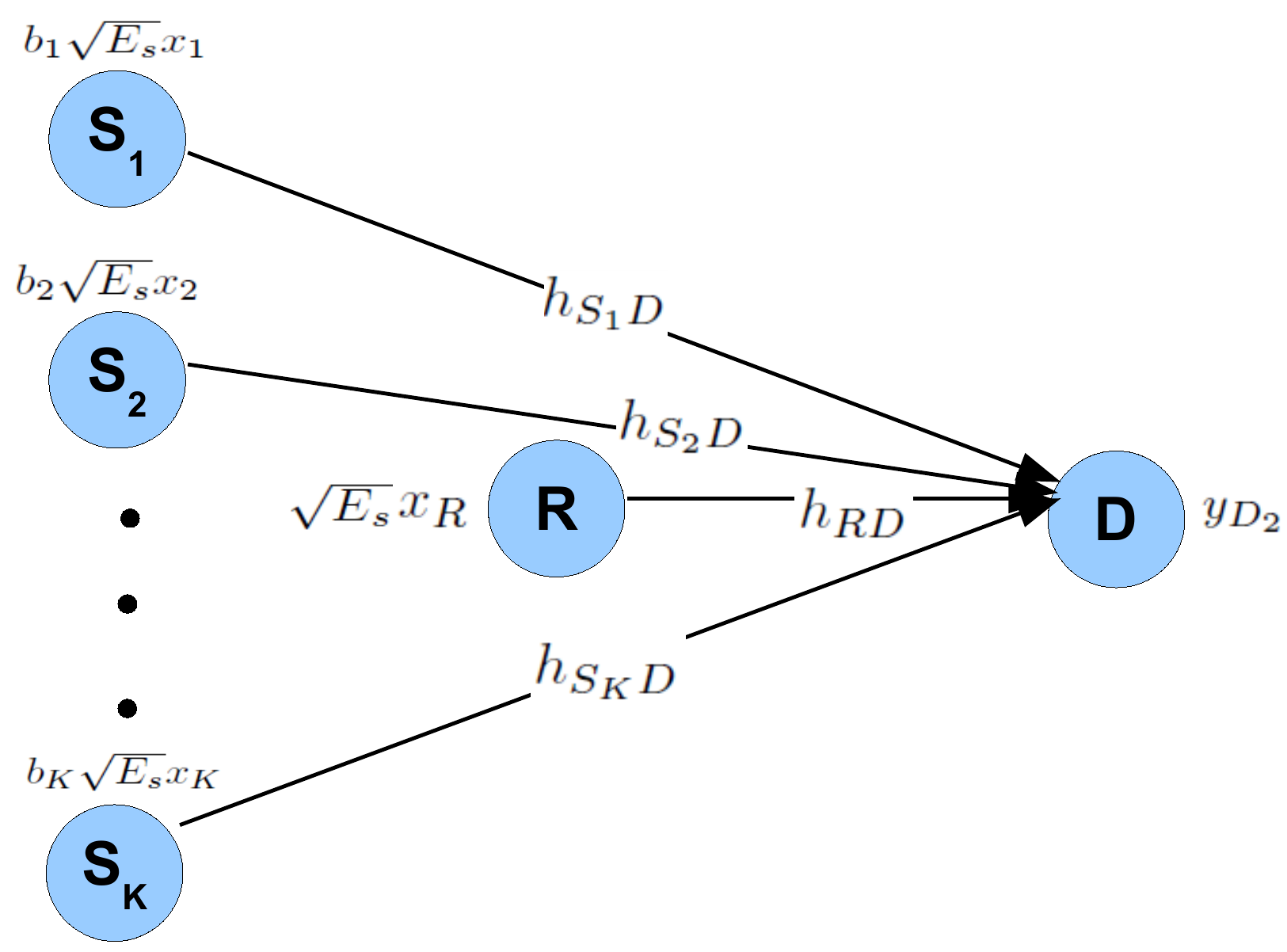}
\label{fig:phase2}
\caption{Phase 2}
\end{subfigure}
\caption{The two user Multiple Access Relay Channel}
\label{fig:MARC}
\vspace{-.5 cm}
\end{figure}

We consider the $K$-user Multiple Access Relay Channel (MARC) shown in Fig. \ref{fig:MARC}. Source nodes $S_1,S_2,\dotso, S_K$ want to transmit messages to the destination node $D$ with the help of the relay node $R.$ All the nodes are assumed to have half-duplex constraint, i.e., the nodes cannot transmit and receive simultaneously in the same frequency band. In a $K$-user MARC, the maximum diversity order obtainable is two, since in addition to the presence of direct links, communication paths exist from the source nodes to the destination node $D$ through the relay node $R.$   
\subsection{Background}
In a fading scenario, Multiple Input Multiple Output (MIMO) antenna systems provide gain in terms of spatial diversity. However, in many practical scenarios, it is difficult to place multiple collocated antennas in a single terminal. An attractive alternative to obtain diversity gain without using multiple antennas, is the utilization of intermediate relay nodes which aid the transmission from the source nodes to the destination nodes. In order to exploit the presence of intermediate relay nodes to obtain diversity gain, the source nodes need to convey their messages to the relay nodes. Due to the superposition nature of the wireless channel, if the source nodes transmit simultaneously in the same frequency band, interference occurs at the relay nodes. A loss of spectral efficiency results, if the source nodes transmit in orthogonal time/frequency slots. A solution to this problem is the use of physical layer network coding, first introduced in \cite{ZhLiLa}, in which the nodes are allowed to transmit simultaneously. Physical layer Network Coding (PNC) has been shown to outperform traditional schemes which involve orthogonal transmissions \cite{ZhLiLa}--\cite{APT1}. So far, most of the works on PNC have mainly focussed only on the two-way relay channel. In our recent work \cite{VvR_2user_MARC}, we proposed a scheme based on PNC for the two user MARC. In this paper, we present the generalization of the scheme proposed in \cite{VvR_2user_MARC} for the $K$-user MARC.

For the MARC, a Complex Field Network Coding (CFNC) scheme was proposed in \cite{WaGi}. The CFNC scheme, like the PNC scheme, avoids the loss of spectral efficiency, by making the source nodes transmit simultaneously. But the major difference between the CFNC scheme and the proposed PNC scheme is that during the relaying phase, the CFNC scheme uses a signal set of size $M^K$ at $R,$ whereas the proposed PNC scheme uses a signal set of size $M,$ where $M$ is the size of the signal set used at the source nodes.
  
As noted in \cite{WaGi}, if the relay node transmits a many-to-one function of the estimates of the messages transmitted by the source nodes and minimum squared Euclidean distance decoder is employed at D, a loss of diversity order results. This problem of loss of diversity order due to error propagation, is encountered in many other wireless scenarios as well and various solutions have been proposed to avoid this problem. Cyclic Redundancy Check bits are used so that the nodes forward only those packets which are decoded correctly \cite{JaHeHuNo}. Some works assume the knowledge of all the instantaneous fade coefficients or error probabilities associated with the intermediate nodes at the destination node, with the decoder at the destination using this knowledge to ensure full diversity \cite{WaCaGiLa},\cite{JuKi}. The CFNC scheme proposed in \cite{WaGi}, uses a scaling factor at the relay node which depends on the instantaneous fade coefficients associated with the links from the source nodes to the relay node, with the scaling factor indicated to the destination using pilot symbols. The proposed scheme does not suffer from the disadvantages of any of the above methods, yet ensures the maximum possible diversity order. This is achieved by means of an efficient choice of the transmission scheme and the use of a novel decoder at the destination $D.$

 For the proposed PNC scheme, making the source nodes also transmit during the relaying phase, combined with a novel decoder ensures the maximum possible diversity order of two. Furthermore, if certain parameters are chosen properly, the proposed decoder for the PNC scheme can be implemented with a decoding complexity order same as that of the CFNC scheme proposed in \cite{WaGi}.
 
For the two-way relay channel, the network coding maps used at the relay node need to form a mathematical structure called Latin Squares, for ensuring unique decodability at the end nodes \cite{NVR}. The structural properties of Latin Squares have been used to obtain the network coding maps in a two-way relay channel \cite{NVR}--\cite{NR_Glcom}. Interestingly, choosing the network coding map used at R to be a $K$-dimensional Latin Hypercube, which is the generalization of the Latin Square to $K$ dimensions, helps towards achieving the maximum diversity order of two for the $K$-user MARC.
  
 The main advantages of the proposed scheme over the CFNC scheme proposed in \cite{WaGi} are summarized below: 
\begin{itemize}
\item 
 In the CFNC scheme, $R$ transmits a complex linear combination of the estimate of the messages transmitted by the source node and the signal set used at $R$ during the relaying phase has $M^K$ points, where $M$ is the size of the signal set used at the source nodes. The minimum distance of the signal set used during the relaying phase vanishes as $K$ increases. In contrast, since the proposed PNC scheme uses a many-to-one map, the signal set used during the relaying phase has only $M$ points. The minimum distance of the signal set used at $R$ is more than that of the CFNC scheme and it remains the same irrespective of the number of source nodes $K.$ Hence the proposed scheme performs better than the CFNC scheme. Simulation results presented for the 3-user and 4-user MARC confirm that the proposed PNC scheme provides a large gain over the CFNC scheme.
 \item
 In the CFNC scheme, $R$ uses a scaling factor which is a function of the instantaneous fade coefficients associated with the links from the source nodes to the relay node, which needs to  be indicated to $D$ using pilot symbols. Since the proposed PNC scheme does not involve any such scaling factor, there is no need of such pilot symbols.
 \end{itemize}
\textbf{\textit{Notations}:}
   Throughout, vectors are denoted by bold lower case letters and matrices are denoted by bold capital letters. The set of complex numbers is denoted by $\mathbb{C}.$ $\mathcal{CN}(0, \sigma ^2)$ denotes a circularly symmetric complex Gaussian random variable with mean zero and variance $\sigma ^2$ and $\mathcal{N}(0, \sigma ^2)$ denotes a real Gaussian random variable with mean zero and variance $\sigma ^2.$ For a matrix $\mathbf{A},$ $\mathbf{A^T}$ and $\mathbf{A^*}$ denotes its transpose and conjugate transpose respectively. For a complex number $x,$ $x^*$ denotes its conjugate and $\vert x \vert$ denotes its absolute value. For a vector $\mathbf{v},$ ${\| \mathbf{v} \|}$ denotes its Euclidean norm. The total transmission energy at a node is assumed to be equal to $E_s$ and all the additive noises are assumed to have a variance equal to $1.$ By SNR, we denote the transmission energy ${E_s}.$ For a signal set $\mathcal{S},$ $\Delta \mathcal{S}$ denotes the difference signal set of $\mathcal{S},$ $\Delta\mathcal{S}=\lbrace x-x' \vert x, x'\in \mathcal{S}\rbrace.$ The all zero matrix of size $n \times n$ is denoted by $\mathbf{O_n}.$ The natural logarithm of $x$ is denoted by $\log(x).$ $\mathbb{E}(X)$ denotes the expectation of $X.$ $Q[.]$ denotes the complementary CDF of the standard Gaussian random variable.

 \vspace{-.2 cm}
 \subsection{Signal Model}
Throughout, a quasi-static fading scenario is assumed with the channel state information available only at the receivers. The source nodes want to transmit a binary vector of length $\lambda$ to the destination node. At each one of the source nodes, the binary vector is mapped onto a point from a $M=2^\lambda$ point signal set denoted by $\mathcal{S}.$ Let $\mu: \mathbb{F}_2^\lambda \rightarrow \mathcal{S}$ denote the mapping from bits to complex symbols used at the source nodes.

The proposed PNC scheme involves two transmission phases: Phase 1 during which 
the source nodes simultaneously transmit and, $R$ and $D$ receive, followed by the Phase 2 during which the source nodes and $R$ transmit to $D$.

\subsubsection*{Phase 1}
 Let $x_i= \mu(s_i)$ $\in \mathcal{S}, i \in \lbrace 1,2\dotso,K \rbrace, s_i \in \mathbb{F}_2^\lambda$ denote the complex symbol the source node $S_i$ wants to convey to $D.$ During Phase 1, the source node $S_i$ transmits a scaled version of $x_i.$ The received signal at $R$ and $D$ during Phase 1 are respectively given by,
{
\begin{align}
\nonumber
&y_R=\sum_{i=1}^K h_{S_i R} \sqrt{E_s} a_i x_i +z_R \text{\;and}\\
\label{eqn_yd1}
&y_{D_1}=\sum_{i=1}^K h_{S_i D} \sqrt{E_s} a_i x_i ++z_{D_1},
\end{align}
}where $a_i \in \mathbb{C}, i \in \lbrace 1,2\dotso,K \rbrace$ are constants and the additive noises $z_R$ and $z_{D_2}$ are assumed to be $\mathcal{CN}(0,1).$ All the fade coefficients are Rayleigh distributed, with the fade coefficient associated with the $S_i$-$R$ link $h_{S_i R} \sim \mathcal{CN}(0,\sigma_{S_i R}^2),$ and the fade coefficient associated with the $S_i$-$D$ link $h_{S_i D} \sim \mathcal{CN}(0,\sigma_{S_i D}^2).$

Based on the received complex number $y_{R},$ the relay node computes the Maximum Likelihood (ML) estimate of $({x}_{1},{x}_{2},\dotso,{x}_{K})$ denoted by $(\hat{x}_{1}^R,\hat{x}_{2}^R,\dotso,\hat{x}_{K}^R),$ i.e.,
{\small $$\displaystyle(\hat{x}_{1}^R,\hat{x}_{2}^R,\dotso,\hat{x}_{K}^R)=\arg\min_{({x}'_{1},{x}'_{2},\dotso,{x}'_{K}) \in \mathcal{S}^K} \vert y_R-\sum_{i=1}^K h_{S_i R} \sqrt{E_s} a_i x'_i\vert.$$}
\subsubsection*{Phase 2}
During Phase 2, the source node $S_i$ transmits a scaled version of $x_i$ and $R$ transmits $x_R=f(\hat{x}_{1}^R,\hat{x}_{2}^R,\dotso,\hat{x}_{K}^R),$ where $f:\mathcal{S}^K \rightarrow \mathcal{S}$ is a many-to-one function. The received signal at $D$ during Phase 2 is given by,  

{\vspace{-.3 cm}
\footnotesize
\begin{align}
\label{eqn_yd2}
y_{D_2}=\sum_{i=1}^K h_{S_iD} \sqrt{E_s} b_i x_i +h_{RD}\sqrt{E_s}x_R+z_{D_2},
\end{align} 
}where $b_i \in \mathbb{C}$ are constants and the additive noise $z_{D_2}$ is assumed to be $\mathcal{CN}(0,1).$ The fade coefficient associated with the $R$-$D$ link $h_{RD}$ is assumed to be $\mathcal{CN}(0,\sigma_{RD}^2).$

For the transmission energy at the source nodes to be equal to $E_s,$ the constants $a_i$ and $b_i$ are chosen such that $\vert a_i\vert^2+\vert b_i \vert ^2 =1, \forall i \in \lbrace 1,2\dotso,K \rbrace.$

From \eqref{eqn_yd1} and \eqref{eqn_yd2}, the received complex numbers at $D$ during the two phases can be written in vector form as, 

{\vspace{-.3 cm}
\scriptsize
\begin{align}
\nonumber
\underbrace{\begin{bmatrix} y_{D_1} & y_{D_2} \end{bmatrix}}_{\mathbf{y_D}}&=  \sqrt{E_s}\underbrace{\begin{bmatrix} h_{S_1 D} & h_{S_2 D} \dotso h_{S_K D} & h_{RD} \end{bmatrix}}_{\mathbf{h}} \underbrace{\begin{bmatrix} a_1 x_1 & b_1 x_1\\a_2 x_2 & b_2 x_2 \\ . & .\\. & .\\. & .\\a_K x_K & b_K x_K \\ 0  & x_R\end{bmatrix}}_{\mathbf{C}(x_1,x_2,\dotso,x_K,x_R)} \\
\label{eqn_vform}
& \hspace{4 cm}+\underbrace{\begin{bmatrix} z_{D_1} & z_{D_2} \end{bmatrix}}_{\mathbf{z_D}}.
\vspace{-.3 cm}
\end{align}}

 The matrix $\mathbf{C}(x_1,x_2,\dotso,x_K,x_R)$ in \eqref{eqn_vform} is referred to as the codeword matrix. The restriction of $\mathbf{C}(x_1,x_2,\dotso,x_K,x_R)$ to the first $K$ rows, denoted by $\mathbf{C_r}(x_1,x_2,\dotso,x_K)$ is referred to as the restricted codeword matrix, i.e., {\footnotesize$\mathbf{C_r}(x_1,x_2,\dotso,x_K)=\begin{bmatrix}a_1 x_1 & a_2 x_2 &\dotso &a_K x_K \\ b_1 x_1 & b_2 x_2& \dotso & b_K x_K \end{bmatrix}^T.$} The difference between any two restricted codeword matrices is referred as the restricted codeword difference matrix, i.e., the restricted codeword difference matrices are of the form {\footnotesize$\mathbf{C_r}(\Delta x_1, \Delta x_2,\dotso,\Delta x_K)=\begin{bmatrix}a_1 \Delta x_1 & a_2 \Delta x_2 &\dotso &a_K \Delta x_K \\ b_1 \Delta x_1 & b_2 \Delta x_2& \dotso & b_K \Delta x_K \end{bmatrix}^T,$} where $\Delta x_i \in \Delta \mathcal{S}, \forall i \in \lbrace 1,2,\dotso,K \rbrace.\vspace{.05 cm}$ 
 
From \eqref{eqn_vform}, the vector $\mathbf{y_D}$ can also be written as, {\footnotesize $$\mathbf{y_D}=\sum_{i=1}^{K} \sqrt{E_s}x_i \mathbf{h}\mathbf{W_i}+\sqrt{E_s}x_R \mathbf{h} \mathbf{W_R}+\mathbf{z_D},$$} \noindent \hspace{-.3 cm} where $\mathbf{W_i}$ is a $(K+1) \times 2$ matrix whose $i^{\text{th}}$ row is given by $[a_i \; b_i]$ and all other entries are zeros. For the $(K+1) \times 2$ matrix $\mathbf{W_R},$ the ${(K+1)}^{\text{th}}$ row is given by $[0\;1]$ and all other entries are zeros. 
The matrices $\mathbf{W_i}, i \in \lbrace 1,2\dotso,K \rbrace$ and $\mathbf{W_R}$ are referred to as the weight matrices. 

The contributions and organization of the paper are as follows: 
A novel decoder for the proposed PNC scheme is presented in Section II A.
In Section II B, it is shown that the decoder presented in Section II A achieves the maximum diversity order of two if the following two conditions are satisfied: (i) the map $f$ used at the relay node forms a $K$-dimensional Latin Hypercube and (ii) the constants $a_i$ and $b_i, i \in \lbrace 1,2\dotso,K \rbrace$ are such that every $2 \times 2$ square submatrix of the restricted codeword difference matrices have rank two when $\Delta x_i$ takes non-zero values.
In Section III, the condition under which the proposed decoder admits fast decoding is obtained. It is shown that when at least one of the weight matrices $\mathbf{W_{i}}$ is Hurwitz-Radon orthogonal  with $\mathbf{W_R},$ the proposed decoder admits fast decoding, with the decoding complexity order same as that of the CFNC scheme proposed in \cite{WaGi}.
Simulation results which show that the proposed PNC scheme provides large gain over the CFNC scheme are presented in Section IV. 
\vspace{-.2 cm}
\section{A Novel Decoder for the Proposed PNC Scheme and its Diversity Analysis}
In the following subsection, a novel decoder for the proposed PNC scheme is presented.
\subsection{A Novel Decoder for the Proposed PNC Scheme}
 When $D$ uses the minimum squared Euclidean distance decoder given by, 

{\vspace{-.2 cm}
\scriptsize
\begin{align*}
\left(\hat{x}_1^D, \hat{x}_2^D, \dotso, \hat{x}_K^D\right)&= \arg \min_{\left(x_1,x_2, \dotso,x_K \right) \in \mathcal{S}^K} \left\lbrace\vert y_{D_1}-\sum_{i=1}^{K}h_{S_iD} \sqrt{E_s} a_i \: x_i\vert^2\right.\\
&\hspace{-1.2 cm}\left.+ \vert y_{D_2} -\sum_{i=1}^{K}h_{S_iD} \sqrt{E_s} b_i\: x_i- h_{RD} \sqrt{E_s} f(x_1,x_2,\dotso,x_K) \vert ^2\right\rbrace,
\end{align*}
\vspace{-.5 cm}}

\noindent a loss of diversity order results, since this decoder does not consider the possibility of decoding errors at the relay node.

Alternatively, we propose a novel decoder given by, 

{\vspace{-.3 cm}
\footnotesize
\begin{align}
\nonumber
\left(\hat{x}_1^D, \hat{x}_2^D, \dotso, \hat{x}_K^D\right)&= \arg \hspace{-.6 cm}\min_{\left(x_1,x_2, \dotso,x_K \right) \in \mathcal{S}^K}\hspace{-.1 cm}  \left \lbrace m_1\left(x_1,x_2,\dotso, x_K\right), \right.\\
\label{eqn_decoder}
&\left.\hspace{.75 cm}\log \left(SNR\right)+m_2\left(x_1,x_2,\dotso, x_K\right)\right \rbrace,
\end{align}}where the metrics $m_1$ and $m_2$ are given in \eqref{metric_m1} and \eqref{metric_m2} respectively, at the top of the next page.

\begin{figure*}
\scriptsize
\begin{align}
\label{metric_m1}
&m_1\left(x_1,x_2,\dotso,x_K\right)=\left\vert y_{D_1}-\sum_{i=1}^{K} h_{S_i D} \sqrt{E_s} a_i \: x_i \right \vert^2+ \left\vert y_{D_2} -\sum_{i=1}^{K}h_{S_i D} \sqrt{E_s} b_i \: x_i- h_{RD} \sqrt{E_s} f(x_1,x_2,\dotso,x_K) \right\vert ^2,\\
\label{metric_m2}
&m_2\left(x_1,x_2,\dotso,x_K\right)=\left\vert y_{D_1}-\sum_{i=1}^{K} h_{S_i D} \sqrt{E_s} a_i \: x_i\right \vert^2+\min_{x_R \neq f\left(x_1,x_2,\dotso, x_K \right), x_R \in \mathcal{S}}\left \lbrace \left\vert y_{D_2} -\sum_{i=1}^{K}h_{S_i D} \sqrt{E_s} b_i \: x_i- h_{RD} \sqrt{E_s} x_R \right\vert ^2\right \rbrace.\\
\hline
\label{metric_m3}
&m_3\left(x_1,x_2,\dotso, x_K\right)=\left\vert y_{D_1}-\sum_{i=1}^{K} h_{S_i D} \sqrt{E_s} a_i \: x_i\right \vert^2 +\min_{x_R \in \mathcal{S}}\left \lbrace \left\vert y_{D_2} -\sum_{i=1}^{K}h_{S_i D} \sqrt{E_s} b_i \: x_i- h_{RD} \sqrt{E_s} x_R \right\vert ^2\right \rbrace.
\end{align}
\hrule
\vspace{-0.4cm}
\end{figure*}
 
The idea behind the choice of this decoder is as follows: 
The optimal ML decoding metric at $D$ is equal to $m_1(x_1,x_2,\dotso,x_K),$ when the relay transmits the correct network-coded symbol.
 The relay transmits a wrong network-coded symbol, independent of $(x_1,x_2,\dotso,x_K),$ if the joint ML estimate at the relay $(\hat{x}_1^R, \hat{x}_2^R,\dotso,\hat{x}_K^R)$ is such  that $x_R=f(\hat{x}_1^R, \hat{x}_2^R,\dotso,\hat{x}_K^R) \neq  f(x_1,x_2, \dotso , x_K).$ Under this condition, the optimal ML decision metric at $D$ is given by $m_2(x_1,x_2, \dotso ,x_K).$
At high SNR, the relay transmits a wrong network-coded symbol with a probability which is proportional to $\frac{1}{SNR}.$ Hence, to the metric $m_2(x_1,x_2, \dotso ,x_K),$ we add a correction factor of $\log(SNR)$ and the minimum of $m_1(x_1,x_2, \dotso ,x_K)$ and $\log(SNR)+m_2(x_1,x_2, \dotso ,x_K)$ is taken to be the decoding metric at $D$.

The CFNC scheme proposed in \cite{WaGi} uses the minimum squared Euclidean distance decoder, which has a decoding complexity of $\mathcal{O}(M^K).$ Since the decoder given in \eqref{eqn_decoder} involves minimization over $K+1$ variables $x_1,x_2, \dotso x_K$ and $x_R,$ it appears as though the decoding complexity order is $\mathcal{O}(M^{K+1}).$ In Section III, it is shown that by properly choosing the constants $a_i$'s and $b_i$'s, the decoding complexity order can be reduced to $\mathcal{O}(M^K)$ which is the same as that of the CFNC scheme.

\subsection{Diversity Analysis of the Proposed Decoder}
The following theorem gives a sufficient condition under which the decoder given in \eqref{eqn_decoder} offers maximum diversity order two.
\begin{theorem}
For the proposed PNC scheme, the decoder given in \eqref{eqn_decoder} offers maximum diversity order two if the following two conditions are satisfied:
\begin{enumerate}
\item
The map $f$ satisfies the condition,

{\vspace{-.2 cm}
\footnotesize
\begin{align}
\nonumber
&f(x_1,x_2,\dotso ,x_{i-1},x_i,x_{i+1}, \dotso, x_K) \\
\label{ex_law}
&\hspace{1.8 cm}\neq f(x_1,x_2,\dotso, x_{i-1}, x'_i,x_{i+1}, \dotso, x_K), \;  \; 
\end{align}}
for $x_i \neq x'_i,$ for all $i \in \lbrace 1,2,\dotso, K \rbrace.$
\item
All $2 \times 2$ submatrices of the restricted codeword difference matrices $\mathbf{C_r}(\Delta x_1, \Delta x_2,\dotso,\Delta x_K)$ have rank two, $\forall \Delta x_1, \Delta x_2, \dotso, \Delta x_K \neq 0.$ 
\end{enumerate} 
\end{theorem}


It is easy to verify that a map $f$ satisfying condition 1) above forms a Latin Hypercube of order $M$ and dimension $K.$
\begin{definition}{\cite{Ki}}
A Latin Hypercube of order $M$ and dimension $K$ is an array of dimension $K,$ with the indices for the $K$ dimensions as well as the entries filled in the array taking values from the symbol set $\lbrace 0,1\dotso ,M-1 \rbrace.$  Every symbol occurs exactly once along each one of the $K$ dimensions.
\end{definition}

The $i^{\text{th}}$ dimension of the Latin Hypercube represents the transmission $x_i$ of the source node $S_i.$ For simplicity, the points of the $M$ point signal set $\mathcal{S}$ are indexed by integers from $0$ to $M-1.$ The entries filled in the Latin Hypercube represent the transmission of the relay node.

\begin{figure}[h]
\scriptsize
\centering
\begin{tabular}{c|c|c|c|c|}
\hline & 0 & 1 & 2 & 3\\
\hline 0 &  0&1  & 2 &3 \\ 
\hline 1 &1  &2 & 3 &0 \\ 
\hline 2 &2  &3  &0  &1 \\ 
\hline 3 & 3 &0  &1&2\\ 
\hline 
\end{tabular}
\caption{An example of Latin Square of order 4}
\label{Latinsq_Example}
\end{figure} 

A Latin Hypercube of dimension 2 is a Latin Square. Fig. \ref{Latinsq_Example} shows an example of a Latin Square of order 4. It can be seen from Fig. \ref{Latinsq_Example} that no two entries repeat in a row as well as a column.

\begin{figure}[htbp]
\centering
\includegraphics[totalheight=1in,width=3.5in]{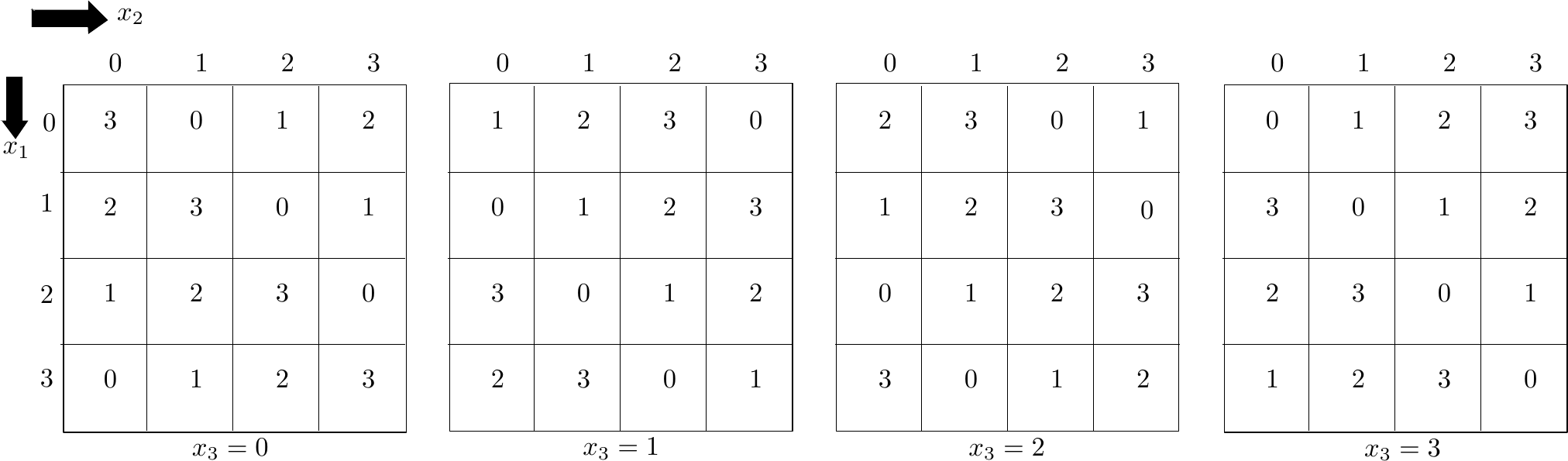}
\caption{An example of Latin Cube of order 4}	
\label{Latin_cube}
\end{figure}

A Latin Hypercube of dimension 3 is a Latin Cube. The three dimensions of a Latin Cube are referred to as rows, columns and pages. Fig. \ref{Latin_cube} shows an example of a Latin Cube of order 4. It can be verified from Fig. \ref{Latin_cube} that in each one of the four pages, no two entries repeat in a row as well as a column. Similarly, for a fixed value of row index and a fixed value of column index, the entries in the four pages are distinct.

\begin{figure}[htbp]
\centering
\includegraphics[totalheight=3.5in,width=3.5in]{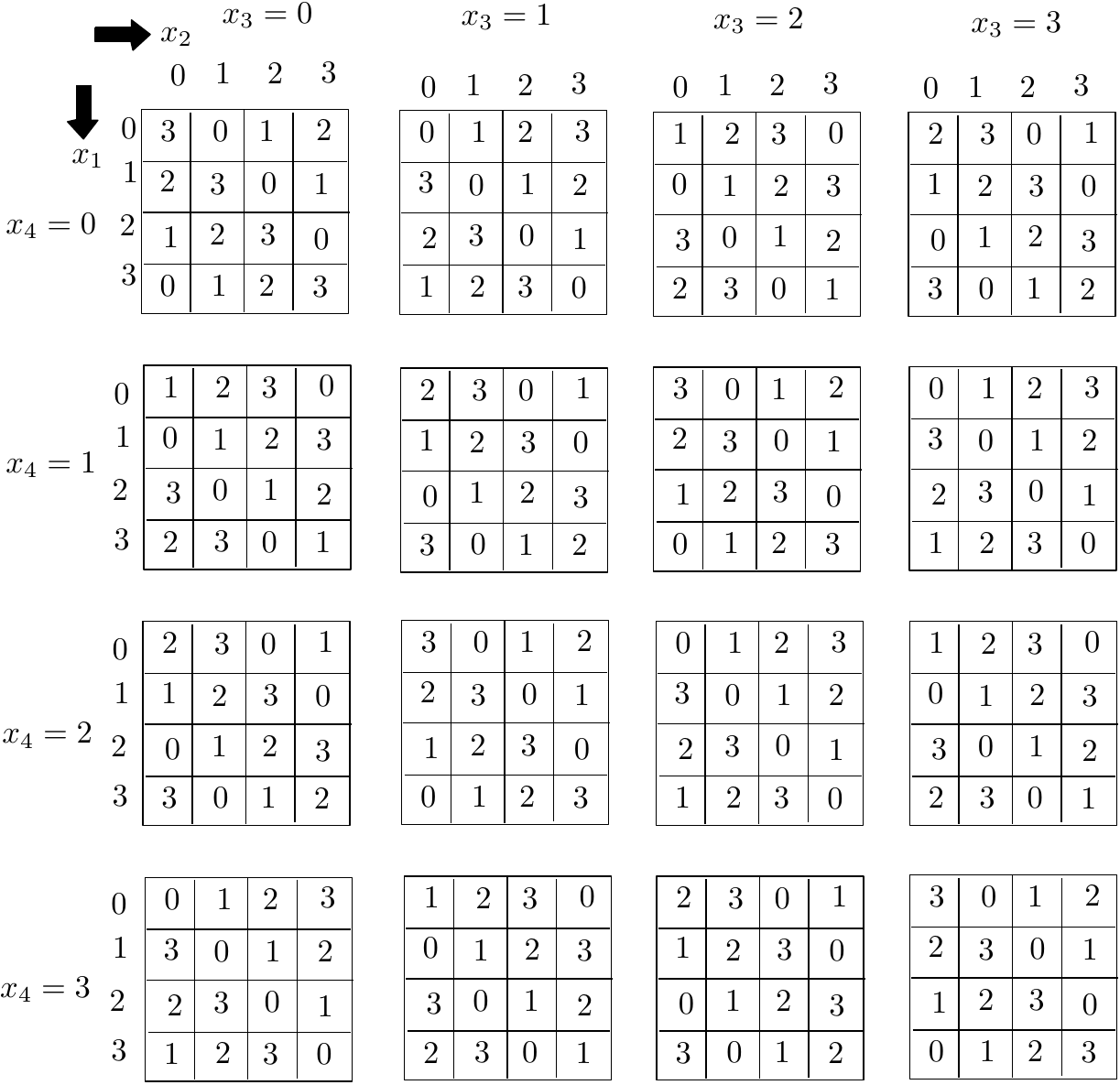}
\caption{An example of Latin Hypercube of dimension 4 and order 4}	
\label{Latin_hypercube_4}
\end{figure}

Fig. \ref{Latin_hypercube_4} shows a Latin Hypercube of dimension 4 and order 4. In Fig. \ref{Latin_hypercube_4} with a $4 \times 4$ square for the which the indices of the third and fourth dimensions are fixed, no two entries repeat in a row as well as a column. Also, when the first three dimensions are fixed, no two entries repeat when the fourth dimension is varied. Similarly, fixing the first, second and fourth dimensions, the four entries obtained when the third dimension is varied are distinct.

For the 3-user MARC, the following example gives a choice of $a_i$'s and $b_i$'s for which Condition 2) given in Theorem 1 is satisfied.   
\begin{example}
For the 3-user MARC, choosing $a_1=1,$ $b_1=0,$ $a_2=\frac{1}{\sqrt{2}},$ $b_2=\frac{1}{\sqrt{2}},$ $a_3=\frac{1}{\sqrt{2}}$ and $b_3=-\frac{1}{\sqrt{2}},$ the restricted codeword difference matrices are of the form $\mathbf{C_r}(\Delta x_1, \Delta x_2, \Delta x_3)= \begin{bmatrix} \Delta x_1 & 0 \\ \frac{1}{\sqrt{2}} \Delta x_2 & \frac{1}{\sqrt{2}} \Delta x_2 \\ \frac{1}{\sqrt{2}}\Delta x_3 & -\frac{1}{\sqrt{2}} \Delta x_3 \end{bmatrix}.$ It can be verified that when $\Delta x_1, \Delta x_2$ and $\Delta x_3$ take non-zero values, the rank of every $2 \times 2$ square submatrix of $\mathbf{C_r}(\Delta x_1, \Delta x_2, \Delta x_3)$ is two and hence condition 2) given in Theorem 1 is satisfied.  
\end{example}

\begin{example}
For the 4-user MARC, choosing $a_1=1,$ $b_1=0,$ $a_2=\frac{1}{\sqrt{2}},$ $b_2=\frac{1}{\sqrt{2}},$ $a_3=\frac{1}{\sqrt{2}},$ $b_3=-\frac{1}{\sqrt{2}},$ $a_4=\frac{j}{\sqrt{2}}$ and $b_4=\frac{1}{\sqrt{2}},$ the restricted codeword difference matrices are of the form $\mathbf{C_r}(\Delta x_1, \Delta x_2, \Delta x_3, \Delta x_4)= \begin{bmatrix} \Delta x_1 & 0 \\ \frac{1}{\sqrt{2}} \Delta x_2 & \frac{1}{\sqrt{2}} \Delta x_2 \\ \frac{1}{\sqrt{2}}\Delta x_3 & -\frac{1}{\sqrt{2}} \Delta x_3\\ \frac{j}{\sqrt{2}}\Delta x_4 & \frac{1}{\sqrt{2}} \Delta x_4 \end{bmatrix}.$ It can be verified that when $\Delta x_1, \Delta x_2, \Delta x_3$ and $\Delta x_4$ take non-zero values, the rank of every $2 \times 2$ square submatrix of $\mathbf{C_r}(\Delta x_1, \Delta x_2, \Delta x_3, \Delta x_4)$ is two and hence condition 2) given in Theorem 1 is satisfied.  
\end{example}

In general, for the $K$-user MARC, there are many possible ways of choosing $a_i$'s and $b_i$'s so that condition 2) given in Theorem 1 is satisfied. Choosing $K$ unit-norm vectors $[a_i\;b_i], i \in \{1,2,\dotso,K\},$ from $\mathbb{C}^2$ such that $[a_i\;b_i]\neq c[a_j\;b_j], c \in \mathbb{C},$ for all $i \neq j$ ensures that 
condition 2) given in Theorem 1 is satisfied. One particular choice of $a_i$'s and $b_i$'s which satisfies the above condition for the $K$-user MARC is given in the next example.
\begin{example}
Consider the set of vectors over $\mathbb{C}^2$ given by {\scriptsize $$V=\{[\cos(\theta) e^{j \phi} \;\sin(\theta)],[-\sin(\theta)  \;\cos(\theta)e^{-j\phi}] :0 < \theta <\frac{\pi}{2}, -\pi \leq \phi <\pi\}.$$} For the  $K$-user MARC, choosing $[a_1 \; b_1]=[0\;0]$ and $[a_{i} \; b_i], i \in \{2,3,\dotso,K\},$ to be any $K-1$ distinct vectors from the set $V$ ensures that condition 2) given in Theorem 1 is satisfied.
\end{example} 
\section{A Fast Decoding Algorithm for the Proposed Decoder}
In this section, it is shown that if the constants $a_i$'s and $b_i$'s are chosen properly, the decoder given in \eqref{eqn_decoder} can be implemented using an efficient algorithm with a complexity order $\mathcal{O}(M^K).$ Note that for the CFNC scheme proposed in \cite{WaGi}, the decoding complexity order at $D$ is $\mathcal{O}(M^K).$ 

Before the algorithm is presented, some notations are introduced.
The points in the signal set $\mathcal{S}$ are denoted by $s_i, 1 \leq i \leq M.$ 

From \eqref{eqn_vform}, the vector $\mathbf{y_D^T}$ can be written as,
\begin{align*}
\scriptsize
\mathbf{y_D^T}\hspace{-.1 cm}=\hspace{-.1 cm}\underbrace{\begin{bmatrix} a_1 h_{S_1D} & a_2 h_{S_2D} & \dotso & a_K h_{S_KD} & 0 \\b_1 h_{S_1D} & b_2 h_{S_2D}&\dotso &b_K h_{S_KD}&  h_{RD} \end{bmatrix}}_{\mathbf{H_{eq}}} \hspace{-.1 cm}\underbrace{\begin{bmatrix} x_1 \\ x_2 \\.\\.\\.\\x_K\\ x_R \end{bmatrix}}_{\mathbf{x}}\hspace{-.2 cm} \sqrt{E_s}+\mathbf{z_D^T}.
\vspace{- .5 cm}
\end{align*}
The matrix $\mathbf{H_{eq}}$ can be decomposed using $\mathbf{QR}$ decomposition as $\mathbf{H_{eq}}=\mathbf{Q}\mathbf{R},$ where $\mathbf{Q}$ is a $2 \times 2$ unitary matrix and $\mathbf{R}=[\mathbf{R_1} \; \mathbf{R_2}]$ is a $2 \times (K+1)$ matrix, with $\mathbf{R_1}$ being upper-triangular of size $2 \times 2$ and $\mathbf{R_2}$ being a $2 \times (K-1) $ matrix. Let $r_{ij}$ denote the $(i,j)^{th}$ entry of $\mathbf{R}.$

Define $\mathbf{\tilde{y}_D}=\mathbf{Q}^T \mathbf{y_D^T}=[\tilde{y}_{D_1} \: \tilde{y}_{D_2}]^T.$
Also, let
{\scriptsize
\begin{align*}
&\phi_1(x_1,x_2,\dotso, x_K)=\vert \tilde{y}_{D_1}-\sum_{i=1}^{K}r_{1i}x_i \sqrt{E_s}\vert^2,\\
&\phi_2(x_1,x_2, \dotso, x_K)=\vert \tilde{y}_{D_2} -\sum_{i=2}^{K} r_{2i}x_i \sqrt{E_s}\\
&\hspace{4 cm}- r_{2(K+1)} f(x_1,x_2, \dotso, x_K) \sqrt{E_s} \vert ^2,\\
&\phi_3(x_2,\dotso, x_K, x_R)=\vert \tilde{y}_{D_2}-\sum_{i=2}^{K} r_{2i}x_i \sqrt{E_s}-r_{2(K+1)}x_R \sqrt{E_s}\vert^2.
\end{align*}}
The following proposition gives a sufficient condition under which {\bf Algorithm 1} below implements the decoder given in \eqref{eqn_decoder}.
\begin{algorithm}
{
\scriptsize
\begin{algorithmic}[1]
\For {$i_2=1$ to $M$}
\For {$i_3=1$ to $M$}
\State .
\State .
\State .
\For {$i_K=1$ to $M$}

    \State $x_2\gets s_{i_2}$
    \State $x_3\gets s_{i_3}$
    \State .
    \State .
    \State .
    \State $x_K \gets s_{i_K}$
    \State \textbf{Find} $\hat{x}_{1}^1=\displaystyle{\arg\min_{x_1 \in \mathcal{S}}\lbrace \phi_1(x_1,x_2, \dotso,x_K) + \phi_2(x_1,x_2, \dotso,x_K)\rbrace}$
     \State \textbf{Find} $\hat{x}_{1}^2=\displaystyle{\arg\min_{x_1 \in \mathcal{S}}\lbrace \phi_1(x_1,x_2, \dotso,x_K)\rbrace}$ 
     \State \textbf{Find} $\hat{x}_R=\displaystyle{\arg\min_{x_R \in \mathcal{S}}\lbrace \phi_3(x_2,\dotso,x_K,x_R)\rbrace}$ 
\If {$\phi_1(\hat{x}_1^1,x_2,\dotso,x_K)+\phi_2(\hat{x}_1^1,{x}_2, \dotso x_K)<\log(SNR)$ $~~~~~~~~~~~~~~~~~~~~~~~~~~~~~~~~~~~~~~~~~~~~~~~~~~~+\phi_1(\hat{x}_1^2,x_2,\dotso, x_K)+\phi_3(x_2,\dotso,x_K,\hat{x}_R)$ $~~~~~~~~~~~~~$}{\\$~~~~~~~~~~~~~~~~m(x_2,\dotso,x_K)=\phi_1(\hat{x}_1^1,x_2,\dotso,x_K)+\phi_2(\hat{x}_1^1,{x}_2,\dotso,x_K)$ \\$~~~~~~~~~~~~~~\hat{x}_1(x_2,\dotso,x_K)=\hat{x}_1^1$}
\Else \\
{$~~~~~~~~~~~~~~~~~~m(x_2,\dotso,x_K)=\phi_1(\hat{x}_1^2,x_2,\dotso,x_K)+\phi_3(x_2,\dotso,x_K,\hat{x}_R)$ \\ $~~~~~~~~~~~~~~~~~~~~~~~~~~~~~~~~~~~~~~~~~~~~~~~~~~~~~~~~~~~~~~~~~~~~~~~~~~~~~~+\log(SNR)$\\$~~~~~~~~~~~~~~\hat{x}_1(x_2,\dotso,x_K)=\hat{x}_1^2$}
\EndIf  
\EndFor
\State .
\State .
\State .
\EndFor  
\EndFor\\
\textbf{Find} ${\displaystyle(\hat{x}_2,\dotso,\hat{x}_K)=\arg \min_{(x_2,\dotso,x_K) \in \mathcal{S}^{K-1}} m(x_2,\dotso x_K)}$\\
$(\hat{x}_1^D,\hat{x}_2^D,\dotso,\hat{x}_K^D)=(\hat{x}_1(\hat{x}_2,\dotso,\hat{x}_K),\hat{x}_2,\dotso,\hat{x}_K)$
\end{algorithmic}}
\caption{Decoding Algorithm used at $D$}
\end{algorithm}
\begin{proposition}
{\bf Algorithm 1}{\footnote{A algorithm exactly similar to {\bf Algorithm 1} can be used with the roles of $S_1$ and $S_i$ interchanged, if $\mathbf{W_i}$ and $\mathbf{W_R}$ are H-R orthogonal.}} implements the decoder in \eqref{eqn_decoder}, if the constants $a_i$'s and $b_i$'s are such that the weight matrices $\mathbf{W_1}$ and $\mathbf{W_R}$ are Hurwitz-Radon (H-R) orthogonal, i.e.,
$\mathbf{W_1} \mathbf{W_R}^{*}+\mathbf{W_R} \mathbf{W_1}^{*}=\mathbf{O_{K+1}}.$ 
\begin{proof}
The decoding metric of the decoder given in \eqref{eqn_decoder} can be written as,

{\vspace{-.3 cm}
\scriptsize
\begin{align}
\nonumber
&\min_{\left(x_1,x_2,\dotso,x_K \right)} \left\lbrace m_1\left(x_1,x_2,\dotso, x_K\right) ,\log\left(SNR\right)+ m_2\left(x_1,x_2,\dotso, x_K \right) \right\rbrace\\
\nonumber
&\hspace{.3 cm}=\min_{\left(x_1,x_2,\dotso, x_K\right)} \lbrace m_1\left(x_1,x_2,\dotso, x_K \right),\\
\nonumber
&\hspace{.4 cm}\log\left(SNR\right)+ m_1\left(x_1,x_2,\dotso, x_K\right),\log\left(SNR\right)+ m_2\left(x_1,x_2,\dotso, x_K\right)\rbrace.\\
\label{decoding_metric}
&\hspace{0.3 cm}=\min_{\left(x_1,x_2,\dotso, x_K\right)} \lbrace m_1\left(x_1,x_2,\dotso, x_K\right), \\
\nonumber
&\hspace{4 cm}\log\left(SNR\right)+ m_3\left(x_1,x_2,\dotso, x_K\right)\rbrace.
\end{align}
}where the metric $m_3(x_1,x_2,\dotso,x_K)$ is given in \eqref{metric_m3}, at the top of the previous to the previous page.
%
We have,
\begin{align}
\nonumber
m_3(x_1,x_2,\dotso,x_K)&= \min_{x_R \in \mathcal{S}} \lbrace \|\mathbf{y_{D}}^T-\mathbf{H_{eq}}\mathbf{x}\sqrt{E_s}\|^2\rbrace\\
\label{metric_m3_second}
&=\min_{x_R \in \mathcal{S}} \lbrace \|\mathbf{\tilde{y}_D}-\mathbf{R}\mathbf{x}\sqrt{E_s}\|^2\rbrace.
\end{align}
Since $\mathbf{R_1}$ is upper triangular, $r_{21}=0.$ Also, the entry $r_{1(K+1)}=0,$ since $\mathbf{W_A}$ and $\mathbf{W_R}$ are H-R orthogonal (follows from Theorem 2, \cite{SrRa}). Hence, from \eqref{metric_m3_second}, it follows that,
\begin{align}
\nonumber
m_3(x_1,x_2,\dotso, x_K)&=  \vert \tilde{y}_{D_1}-\sum_{i=1}^K r_{1i} x_i \sqrt{E_s}\vert^2\\
&\hspace{-3.5 cm}+ \min_{x_R \in \mathcal{S}}\left \lbrace\vert \tilde{y}_{D_2}-\sum_{i=2}^K r_{2i}x_i \sqrt{E_s}-r_{2(K+1)}x_R \sqrt{E_s}\vert^2\right\rbrace.
\end{align}
Hence, 
\begin{align*}
\nonumber
\min_{x_1 \in \mathcal{S}} m_3(x_1,x_2,\dotso,x_K)&= \min_{x_1 \in \mathcal{S}}\phi_1(x_1,x_2,\dotso,x_K)\\
&+\min_{x_R \in \mathcal{S}}\phi_3(x_2,\dotso,x_K,x_R).
\end{align*}

From \eqref{decoding_metric}, the decoding metric can be written as,
\begin{align*}
&\min_{(x_2,\dotso,x_K) \in \mathcal{S}^{K-1}}\left\lbrace \min_{x_1 \in \mathcal{S}}\left\{\phi_1\left(x_1,x_2,\dotso,x_K \right)\right.\right.\\
&\hspace{5 cm}\left.\left.+\phi_2\left(x_1,x_2,\dotso,x_K\right)\right\}\right., \\
&\left.\hspace{1 cm}\min_{x_1 \in \mathcal{S}}\phi_1\left(x_1,x_2,\dotso,x_K\right)\right.\\
&\left.\hspace{2.3 cm}+\min_{x_R \in \mathcal{S}}\phi_3\left(x_2,\dotso,x_K, x_R\right)+\log\left(SNR\right)\right\rbrace.
\end{align*}
\end{proof}
\end{proposition}

In {\bf Algorithm 1}, inside the $K-1$ nested {\bf for} loops, the values of $x_2, x_3 \dotso, x_K$ are fixed and the operations in lines 13, 14 and 15 involve a complexity order $\mathcal{O}(M).$ The operations from line 16 to line 23 involve constant complexity, independent of $M.$ Hence the complexity order for executing the nested {\bf for} loops from line 1 to line 29 is $\mathcal{O}(M^K).$ The operation in line 30 involves a complexity order $\mathcal{O}(M^{K-1}).$ Hence the overall complexity order of {\bf Algorithm 1} is $\mathcal{O}(M^K),$ which is the same as that of the CFNC scheme proposed in \cite{WaGi}. 
\begin{example}
Continuing with Example 1, for the 3-user MARC, when $a_1=1,$ $b_1=0,$ $a_2=\frac{1}{\sqrt{2}},$ $b_2=\frac{1}{\sqrt{2}}$ and $a_3=\frac{1}{\sqrt{2}},$ and $b_3=-\frac{1}{\sqrt{2}},$ the weight matrices are given by, {\footnotesize$\mathbf{W_1}=\begin{bmatrix}1 & 0 & 0& 0\\0 & 0&0 & 0\end{bmatrix}^T,$} {\footnotesize$\mathbf{W_2}=\begin{bmatrix}0 &  \frac{1}{\sqrt{2}} & 0&0\\0& \frac{1}{\sqrt{2}}&0 & 0 \end{bmatrix}^T,$}  {\footnotesize$\mathbf{W_3}=\begin{bmatrix}0 & 0& \frac{1}{\sqrt{2}} &0\\0&0& -\frac{1}{\sqrt{2}}& 0\end{bmatrix}^T$} and {\footnotesize$\mathbf{W_R}=\begin{bmatrix}0 & 0&0 & 0\\0&0&0 & 1\end{bmatrix}^T.$} It can be verified that the matrices $\mathbf{W_1}$ and $\mathbf{W_R}$ are H-R orthogonal, i.e., $\mathbf{W_1} \mathbf{W_R}^{*}+\mathbf{W_R} \mathbf{W_1}^{*}=\mathbf{O_{4}}.$ Hence, for this case, {\bf Algorithm 1} can be used to implement the decoder given in \eqref{eqn_decoder}.
\end{example}
\vspace{-.2 cm}
\section{Simulation Results}
Simulation results presented in this section compare the performance of the proposed PNC scheme with the CFNC scheme proposed in \cite{WaGi}, for the 3-user and 4-user MARC. In the simulation results presented for 3-user MARC, the values of the constants $a_i$'s and $b_i$'s are chosen to be the ones in Example 1 with 4-PSK signal set is used at the nodes and the Latin Cube given in Fig. \ref{Latin_cube} is used as the network coding map at the relay node.  For the 4-user MARC, the values of the constants $a_i$'s and $b_i$'s are chosen to be the ones in Example 2 with 4-PSK signal set is used at the nodes and the Latin Hypercube of dimension 4 given in Fig. \ref{Latin_hypercube_4} is used as the network coding map at the relay node. 

\begin{figure}[htbp]
\centering
\vspace{-.75 cm}
\includegraphics[totalheight=2.8in,width=3.85in]{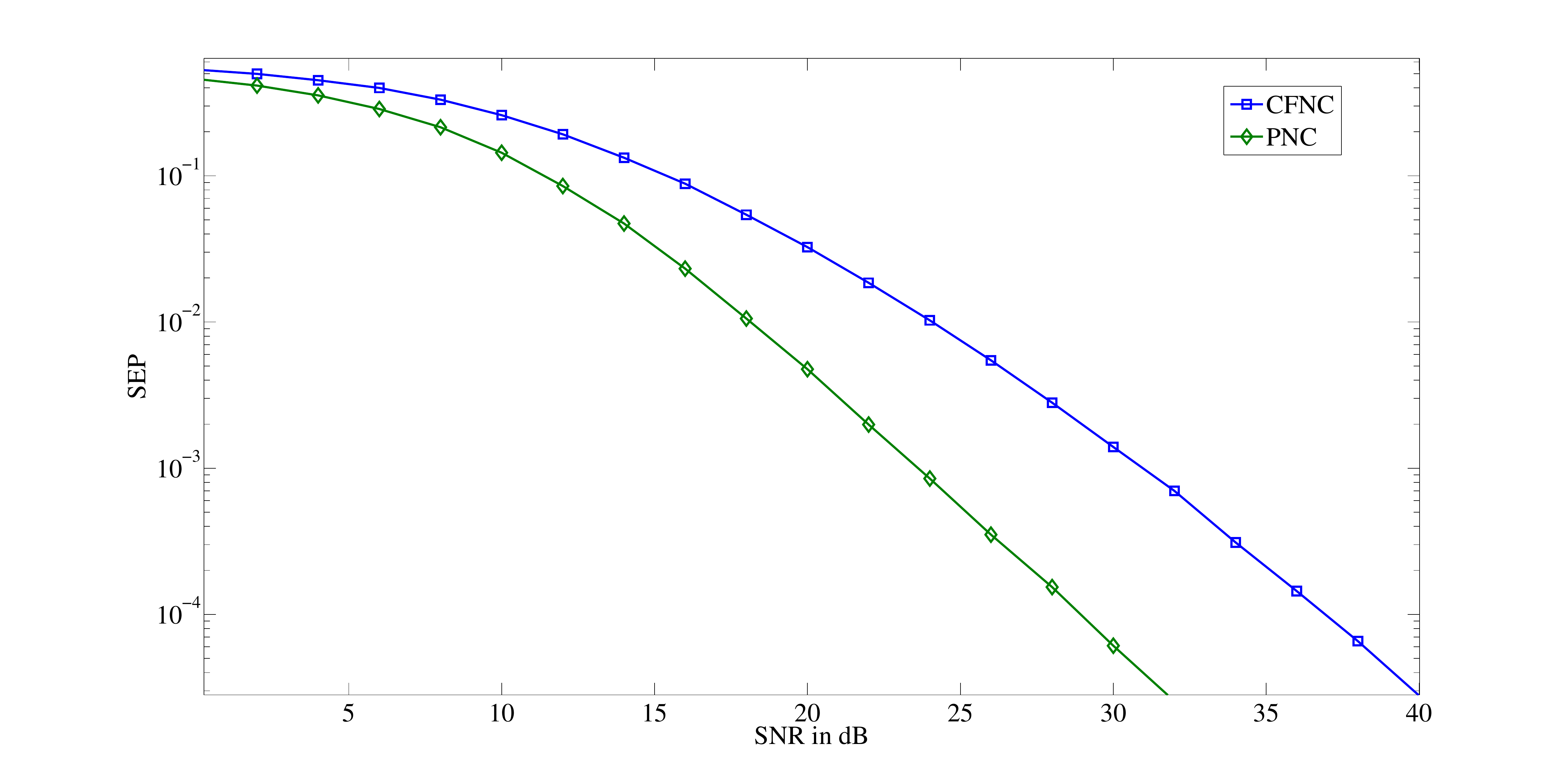}
\vspace{-.75 cm}
\caption{SNR vs SEP plots for 4-PSK signal set for {\small $\sigma_{S_iR}^2=\sigma_{S_iD}^2=\sigma_{RD}^2=0$ dB} for the 3-user MARC with 4-PSK signal set.}	
\label{fig:plots_4psk_equal}
\end{figure}

\begin{figure}[htbp]
\centering
\vspace{-.6 cm}
\includegraphics[totalheight=2.8in,width=3.85in]{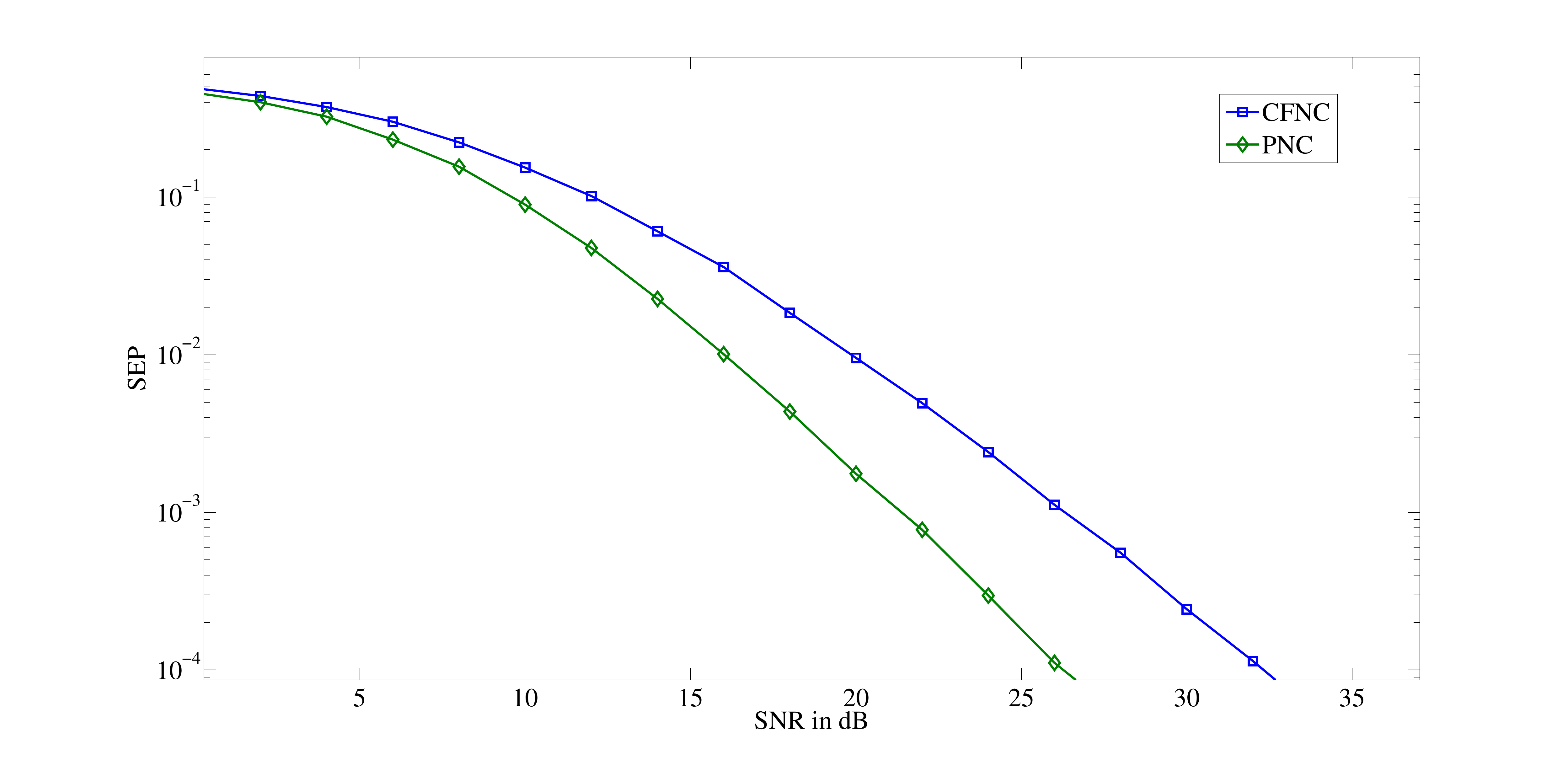}
\vspace{-.75 cm}
\caption{SNR vs SEP plots for {\small $\sigma_{S_iR}^2=10$ dB, $\sigma_{S_iD}^2=\sigma_{BD}^2=\sigma_{RD}^2=0$ dB} for the 3-user MARC with 4-PSK signal set.}	
\label{fig:plots_4psk_sr_strong}
\vspace{-.4 cm}	
\end{figure}

\begin{figure}[htbp]
\centering
\includegraphics[totalheight=2.8in,width=3.85in]{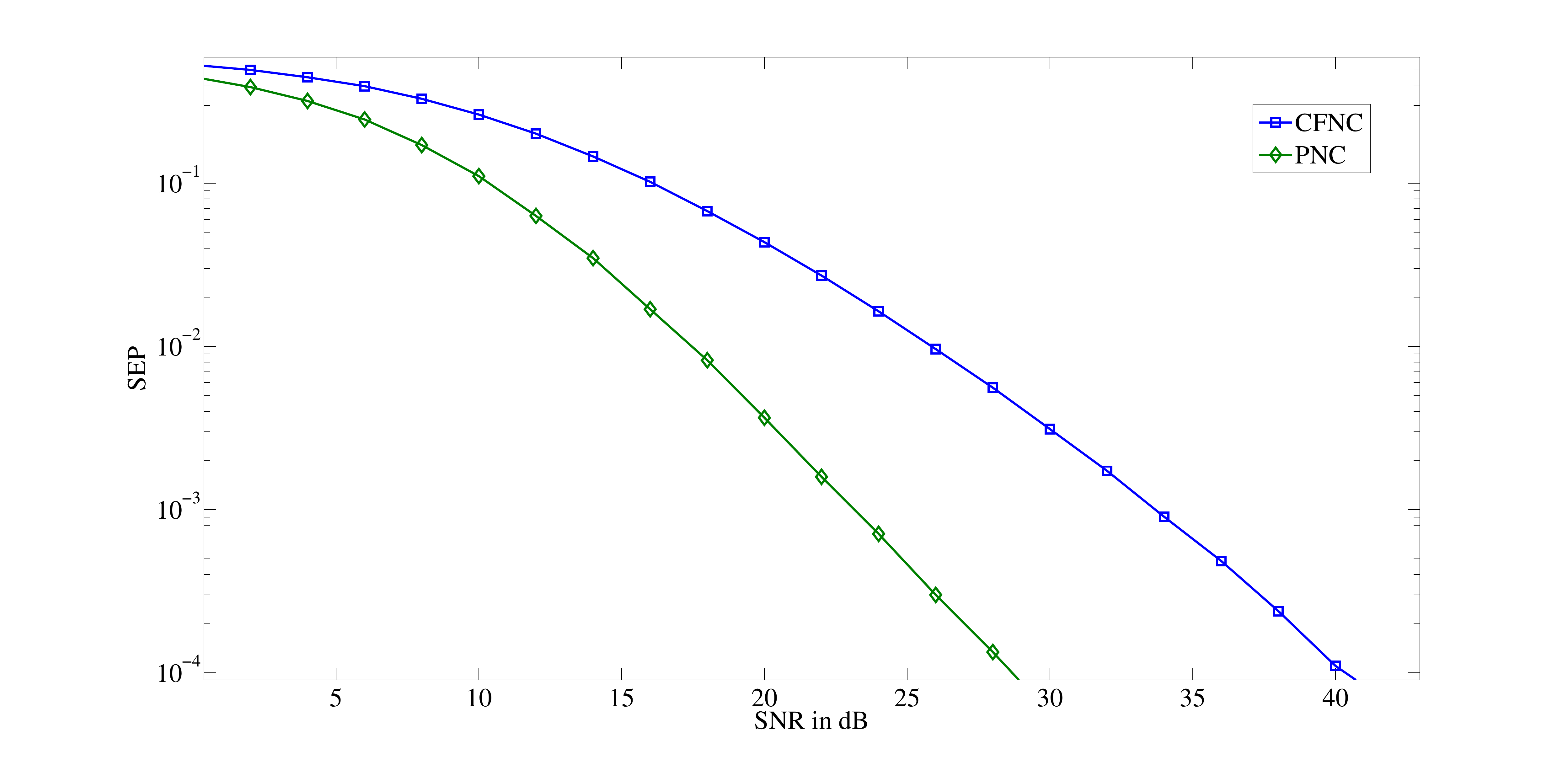}
\vspace{-.75 cm}
\caption{SNR vs SEP plots for 4-PSK signal set for {\small$\sigma_{S_iR}^2=\sigma_{S_iD}^2=0$ dB, $\sigma_{RD}^2=10$ dB} for the 3-user MARC with 4-PSK signal set.}	
\label{fig:plots_4psk_rd_strong}
\vspace{-.7 cm}	
\end{figure}

 For the 3-user MARC, for the case when the variances of all the fading links are 0 dB, the SNR Vs. Symbol Error Probability (SEP) plots are shown in Fig. \ref{fig:plots_4psk_equal}. It can be seen from Fig. \ref{fig:plots_4psk_equal} that the PNC scheme performs better than the CFNC scheme and offers a large gain of 8 dB, when the SEP is $10^{-4}.$ Fig. \ref{fig:plots_4psk_sr_strong} shows a similar plot for the case when  $\sigma_{S_iR}^2=10$ dB  and $\sigma_{S_i D}^2=\sigma_{RD}^2=0$ dB, where $i \in \lbrace 1,2, 3\rbrace.$ It can be seen from Fig. \ref{fig:plots_4psk_sr_strong} that for this case, the PNC scheme offers a gain of nearly 6 dB, when the SEP is $10^{-4}.$ Fig. \ref{fig:plots_4psk_rd_strong} shows the plots for the case when the $R$-$D$ link is stronger than all other links, i.e, $\sigma_{S_iR}^2=\sigma_{S_iD}^2=\sigma_{BD}^2=0 \text{ dB}, i \in \lbrace 1,2,3\rbrace$ and $\sigma_{RD}^2=10$ dB. For this case, the PNC scheme offers a large gain of about 12 dB, when the SEP is $10^{-4}.$ Also, it can be verified from the plots that the proposed decoder for the PNC scheme offers the maximum possible diversity order of two.
 
\begin{figure}[htbp]
\centering
\includegraphics[totalheight=2.8in,width=3.85in]{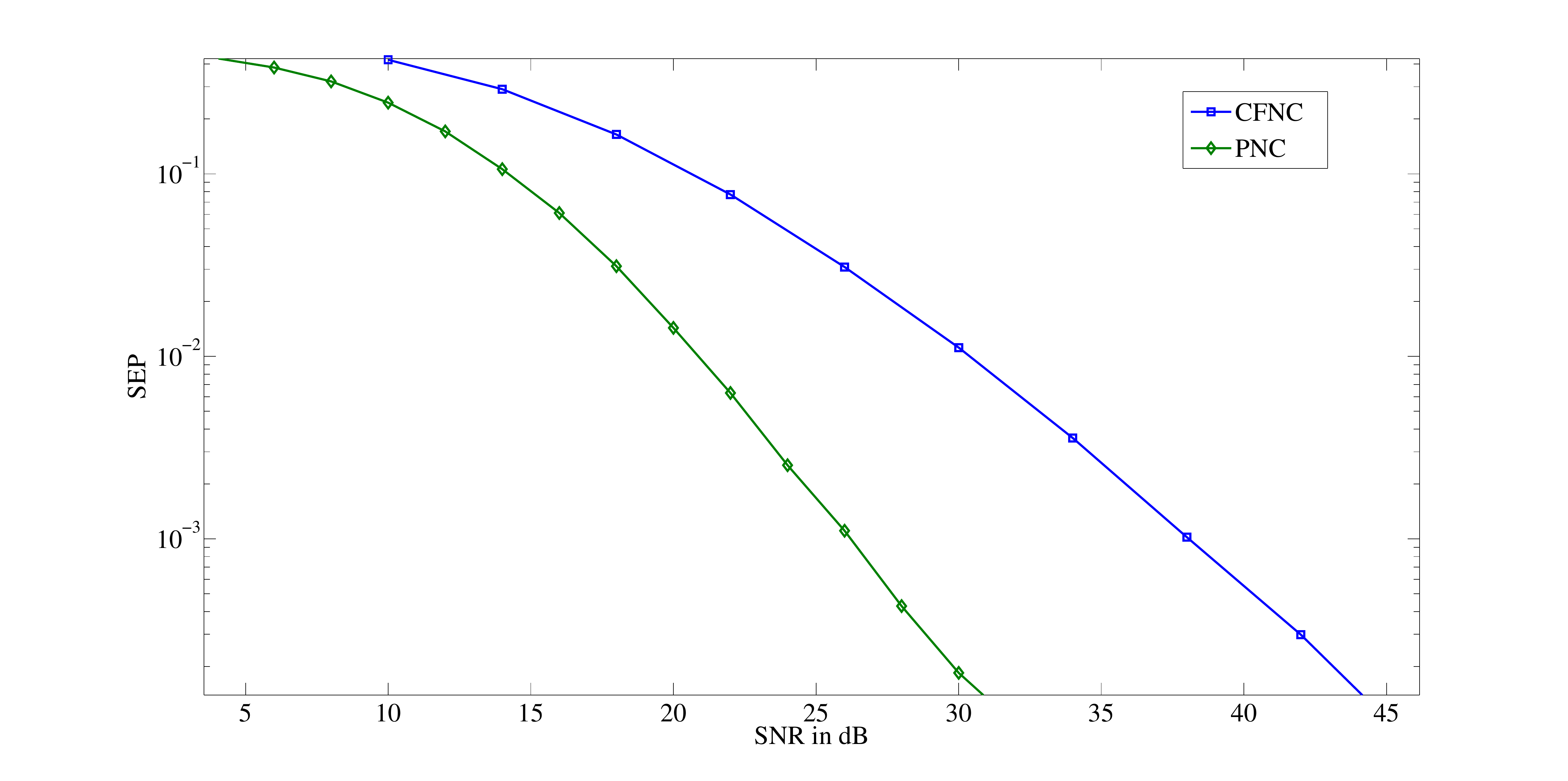}
\vspace{-.75 cm}
\caption{SNR vs SEP plots for 4-PSK signal set for {\small $\sigma_{S_iR}^2=\sigma_{S_iD}^2=\sigma_{RD}^2=0$ dB} for the 4-user MARC with 4-PSK signal set}	
\label{fig:plots_4psk_4source_equal}
\end{figure}

\begin{figure}[htbp]
\centering
\includegraphics[totalheight=2.8in,width=3.85in]{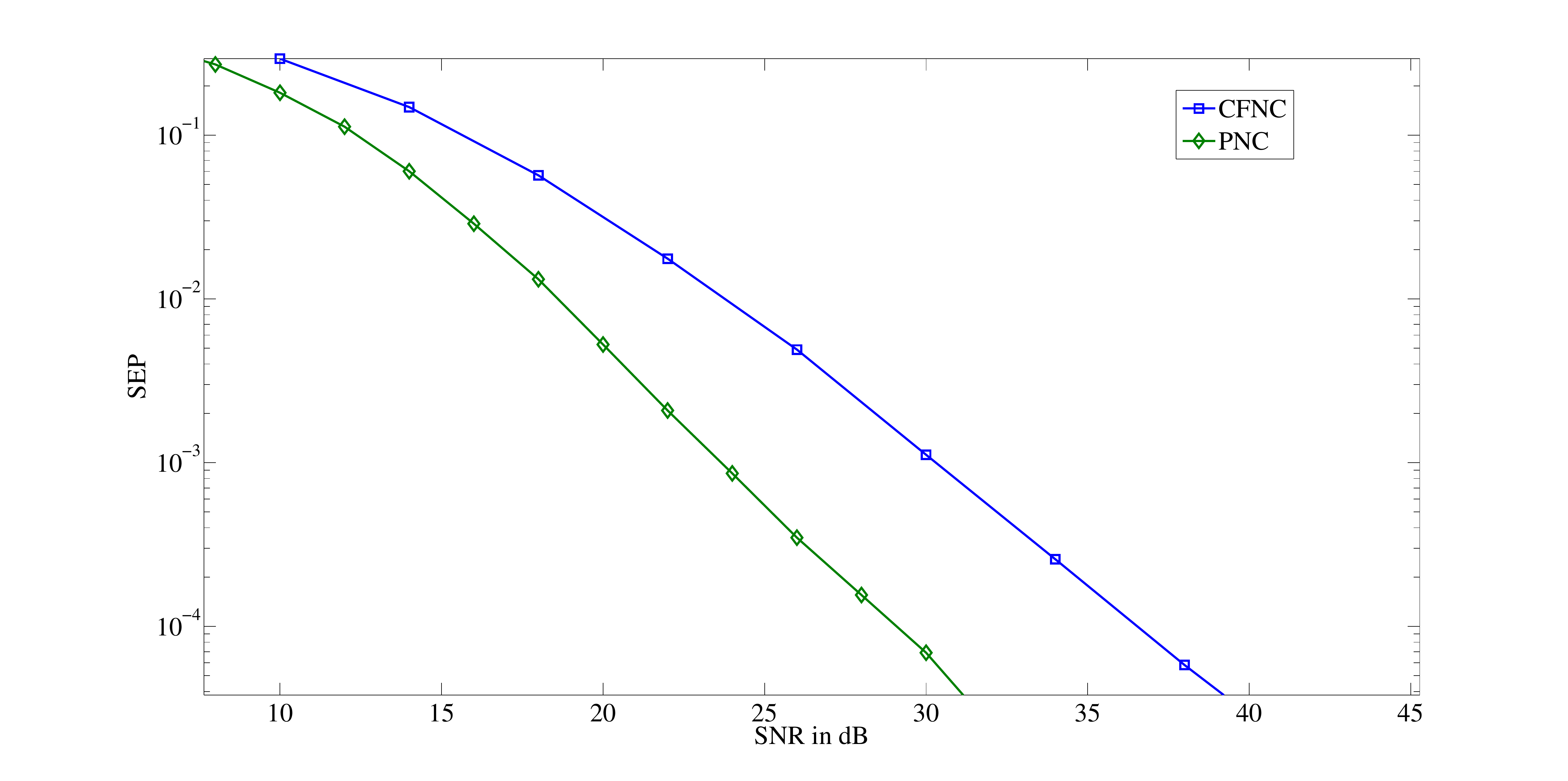}
\vspace{-.75 cm}
\caption{SNR vs SEP plots for 4-PSK signal set for {\small $\sigma_{S_iR}^2=10$ dB, $\sigma_{S_iD}^2=\sigma_{BD}^2=\sigma_{RD}^2=0$ dB} for the 4-user MARC with 4-PSK signal set.}	
\label{fig:plots_4psk_4source_sr_strong}
\end{figure}

\begin{figure}[htbp]
\centering
\vspace{-.5 cm}
\includegraphics[totalheight=2.8in,width=3.85in]{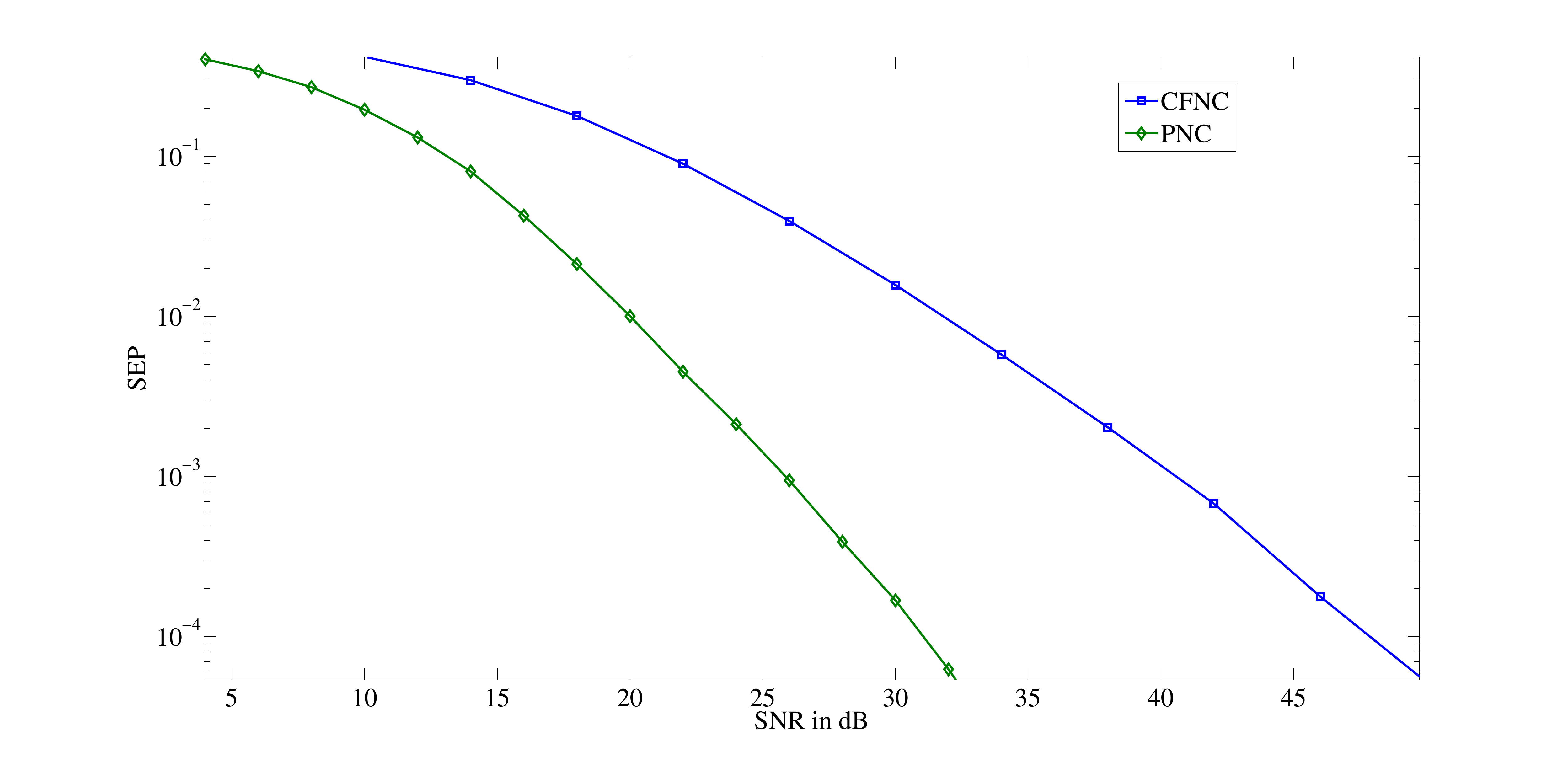}
\vspace{-.75 cm}
\caption{SNR vs SEP plots for 4-PSK signal set for {\small$\sigma_{S_iR}^2=\sigma_{S_iD}^2=0$ dB, $\sigma_{RD}^2=10$ dB} for the 4-user MARC with 4-PSK signal set}	
\label{fig:plots_4psk_4source_rd_strong}
\vspace{-.7 cm}	
\end{figure}

 Fig. \ref{fig:plots_4psk_4source_equal}, Fig. \ref{fig:plots_4psk_4source_sr_strong} and Fig. \ref{fig:plots_4psk_4source_rd_strong} show similar plots for the 4-user MARC with 4-PSK signal set. When the variances of all the fading links are 0 dB, from Fig. \ref{fig:plots_4psk_4source_equal}, it can be seen that the proposed PNC scheme offers a gain of 13 dB, when the SEP is $10^{-4}.$ For the case when  $\sigma_{S_iR}^2=10,i \in \lbrace 1,2, 3\rbrace,$ and all other variances are $0$ dB, from Fig. \ref{fig:plots_4psk_4source_sr_strong} it can be seen that the PNC scheme offers a gain of nearly 7.5 dB, when the SEP is $10^{-4}.$ For the case when the link from $R$ to $D$ is stronger the other links by 10 dB,  it can be seen from Fig. \ref{fig:plots_4psk_4source_rd_strong} that the PNC scheme offers an advantage of 17 dB over the CFNC scheme.
\section{Discussion}
A physical layer network coding scheme was proposed for the $K$-user Multiple Access Relay Channel. For the proposed scheme, a novel decoder was presented and it was shown that the decoder offers the maximum possible diversity order of two if the network coding map used at the relay forms a $K$-dimensional Latin Hypercube and every $2 \times 2$ submatrix of a restricted codeword difference matrix $\mathbf{C_r}(\Delta x_1,x_2,\dotso,\Delta x_K),$ $\Delta x_i \neq 0, \forall i \in \lbrace 1,2,\dotso,K\rbrace,$ has a rank two. Also, it was shown that the proposed decoder can be implemented using a fast decoding algorithm, if a weight matrix $\mathbf{W_i}$ is Hurwitz-Radon orthogonal with $\mathbf{W_R},$ for some $i \in \lbrace1,2\dotso ,K \rbrace.$  The problem of finding the constants $a_i$'s and $b_i$'s which minimize the error probability in addition to ensuring maximum diversity order remains open. Extension of the proposed scheme for the case when there are multiple relay nodes is a possible direction for future work.

\section*{APPENDIX - Proof of Theorem 1}
Let $H$ denote a particular realization of the fade coefficients. Throughout the proof, the subscript $H$ in a probability expression indicates conditioning on the fade coefficients. For simplicity of notation, it is assumed that the variances of all the fading coefficients are one, but the result holds for other values as well.

Let $E$ denote an error event that the transmitted message $K$-tuple $(x_1,x_2,\dotso,x_K)$ is wrongly decoded at D. 

The probability of $E$ conditioned on $H$ given in \eqref{pe_H}, can be upper bounded as in \eqref{pe_H_ub} (eqns. \eqref{pe_H} and \eqref{pe_H_ub} are shown at the next page). $P_{H}\left \lbrace f(\hat{x}_1^R, \hat{x}_2^R,\dotso,\hat{x}_K^R) = f\left(x_1,x_2,\dotso,x_K\right)\right \rbrace$ and $P_{H}\left \lbrace f(\hat{x}_1^R, \hat{x}_2^R,\dotso,\hat{x}_K^R) \neq f\left(x_1,x_2,\dotso,x_K\right)\right \rbrace$ respectively denote the probabilities that $R$ transmits the correct and wrong network coded symbol during Phase 2, for a given $H.$ Also, the probability {\small$P_{H}\left \lbrace(\hat{x}_1^D , \hat{x}_2^D,\dotso, \hat{x}_K^D) \neq \left(x_1,x_2,\dotso,x_K\right)\right.$ $\left.\bigm\vert f(\hat{x}_1^R, \hat{x}_2^R,\dotso, \hat{x}_K^R)= f\left(x_1,x_2,\dotso,x_K\right)\right\rbrace$} and the probability  {\small$P_{H}\left \lbrace(\hat{x}_1^D , \hat{x}_2^D,\dotso,\hat{x}_K^D) \neq \left(x_1,x_2,\dotso,x_K\right)\right.$ $\left.\bigm\vert f(\hat{x}_1^R, \hat{x}_2^R,\dotso,\hat{x}_K^R)\neq f\left(x_1,x_2,\dotso,x_K\right)\right\rbrace$} in \eqref{pe_H} respectively denote the probabilities of $E$ given that $R$ transmitted the correct and wrong network coded symbol for a given $H.$  $P_{H}\lbrace E \rbrace$ can be upper bounded as in \eqref{pe_H_ub_equal}, where $P_{H}\left \lbrace f(\hat{x}_1^R, \hat{x}_2^R,\dotso,\hat{x_K}^R) = x'_R\right \rbrace$ denotes the probability that the network coded symbol transmitted by $R$ is $x'_R \neq f(x_1,x_2,\dotso,x_K)$ and the probability $ P_{H}\left \lbrace(\hat{x}_1^D , \hat{x}_2^D,\dotso, \hat{x}_K^D) \neq \left(x_1,x_2,\dotso,x_K\right)\right.$ $\left. \bigm\vert f(\hat{x}_1^R, \hat{x}_1^R,\dotso,\hat{x}_K^R)= x'_R\right\rbrace$ is the probability of $E$ given that R transmits $x'_R,$ for a given $H.$ Taking expectation of the terms in \eqref{pe_H_ub_equal} w.r.t $H,$ we get \eqref{pe_H_ub_equal_avg}.

\begin{figure*}
\scriptsize
\begin{align}
\nonumber
P_{H}\lbrace E \rbrace &= P_{H}\left \lbrace(\hat{x}_1^D , \hat{x}_2^D,\dotso,\hat{x}_K^D) \neq \left(x_1,x_2,\dotso,x_K\right) \bigm\vert f(\hat{x}_1^R, \hat{x}_2^R,\dotso,\hat{x}_K^R) = f\left(x_1,x_2,\dotso,x_K \right)\right\rbrace P_{H}\left \lbrace f(\hat{x}_1^R, \hat{x}_2^R,\dotso,\hat{x}_K^R) = f\left(x_1,x_2,\dotso, x_K\right)\right \rbrace\\
\label{pe_H}
&\hspace{0.5 cm}+ P_{H}\left \lbrace(\hat{x}_1^D , \hat{x}_2^D,\dotso,\hat{x}_K^D) \neq \left(x_1,x_2,\dotso, x_K\right) \bigm\vert f(\hat{x}_1^R, \hat{x}_2^R,\dotso,\hat{x}_K^R) \neq f\left(x_1,x_2,\dotso, x_K\right)\right\rbrace P_{H}\left \lbrace f(\hat{x}_1^R, \hat{x}_2^R,\dotso,\hat{x}_K^R) \neq f\left(x_1,x_2,\dotso, x_K\right)\right \rbrace\\
\nonumber
&\leq P_{H}\left \lbrace(\hat{x}_1^D , \hat{x}_2^D,\dotso,\hat{x}_K^D) \neq \left(x_1,x_2,\dotso, x_K\right) \bigm\vert f(\hat{x}_1^R, \hat{x}_2^R,\dotso,\hat{x}_K^R)= f\left(x_1,x_2,\dotso, x_K\right)\right\rbrace \\
\label{pe_H_ub}
&\hspace{1 cm}+ P_{H}\left \lbrace(\hat{x}_1^D , \hat{x}_2^D,\dotso,\hat{x}_K^D) \neq \left(x_1,x_2,\dotso, x_K\right) \bigm\vert f(\hat{x}_1^R, \hat{x}_2^R,\dotso,\hat{x}_K^R) \neq f\left(x_1,x_2,\dotso, x_K\right)\right\rbrace P_{H}\left \lbrace f(\hat{x}_1^R, \hat{x}_2^R,\dotso,\hat{x}_K^R) \neq f\left(x_1,x_2,\dotso, x_K\right)\right \rbrace\\
\nonumber
& \leq  P_{H}\left \lbrace(\hat{x}_1^D , \hat{x}_2^D,\dotso,\hat{x}_K^D) \neq \left(x_1,x_2,\dotso, x_K\right) \bigm\vert f(\hat{x}_1^R, \hat{x}_2^R,\dotso,\hat{x}_K^R)= f\left(x_1,x_2,\dotso, x_K\right)\right\rbrace \\
\label{pe_H_ub_equal}
&\hspace{1 cm}+ \sum_{\substack{{f(\hat{x}_1^R, \hat{x}_2^R,\dotso,\hat{x}_K^R) = x'_R,}\\{x'_R \neq f(x_1,x_2,\dotso, x_K)}}}P_{H}\left \lbrace(\hat{x}_1^D , \hat{x}_2^D,\dotso,\hat{x}_K^D) \neq \left(x_1,x_2,\dotso, x_K\right) \bigm\vert f(\hat{x}_1^R, \hat{x}_2^R,\dotso,\hat{x}_K^R)= x'_R\right\rbrace P_{H}\left \lbrace f(\hat{x}_1^R, \hat{x}_2^R,\dotso,\hat{x}_K^R) = x'_R\right \rbrace\\
\nonumber
P\lbrace E \rbrace &\leq P\left \lbrace(\hat{x}_1^D , \hat{x}_2^D,\dotso,\hat{x}_K^D) \neq \left(x_1,x_2,\dotso, x_K\right) \bigm\vert f(\hat{x}_1^R, \hat{x}_2^R,\dotso,\hat{x}_K^R)= f\left(x_1,x_2,\dotso, x_K\right)\right\rbrace \\
\label{pe_H_ub_equal_avg}
&\hspace{1 cm}+ \hspace{-.5 cm}\sum_{\substack{{f(\hat{x}_1^R, \hat{x}_2^R,\dotso,\hat{x}_K^R) = x'_R,}\\{x'_R \neq f(x_1,x_2,\dotso, x_K)}}}\hspace{-.5 cm}P\left \lbrace(\hat{x}_1^D , \hat{x}_2^D,\dotso,\hat{x}_K^D) \neq \left(x_1,x_2,\dotso, x_K\right) \bigm\vert f(\hat{x}_1^R, \hat{x}_2^R,\dotso,\hat{x}_K^R)= x'_R\right\rbrace P\left \lbrace f(\hat{x}_1^R, \hat{x}_2^R,\dotso,\hat{x}_K^R) = x'_R\right \rbrace.
\end{align}
\hrule
\end{figure*}
The rest of the proof of Theorem 1 is presented in two parts as Lemma 1 and Lemma 2. In Lemma 1, it is shown that the probability $P\left \lbrace(\hat{x}_1^D , \hat{x}_2^D,\dotso, \hat{x}_K^D) \neq \left(x_1,x_2,\dotso,x_K\right)\right.$ $\left.\bigm\vert f(\hat{x}_1^R, \hat{x}_2^R,\dotso,\hat{x}_K^R)= f\left(x_1,x_2,\dotso,x_K\right)\right\rbrace $ has a diversity order two. Lemma 2 shows that the probability $P\left \lbrace(\hat{x}_1^D , \hat{x}_2^D,\hat{x}_K^D) \neq \left(x_1,x_2,\dotso,x_K\right)\right.$ $\left.\bigm\vert f(\hat{x}_1^R, \hat{x}_2^R,\dotso,\hat{x}_K^R)= x'_R\right\rbrace$ has a diversity order one. Since $P\left \lbrace f(\hat{x}_1^R, \hat{x}_2^R,\dotso,\hat{x}_K^R) = x'_R\right \rbrace$ has a diversity order one, Lemma 1 and Lemma 2 together imply that $P\left \lbrace E \right\rbrace$ has a diversity order two.

\begin{lemma}
 When the two conditions in the statement of Therorem 1 are satisfied, the probability $P\left \lbrace(\hat{x}_1^D , \hat{x}_2^D,\dotso\hat{x}_K^D) \neq \left(x_1,x_2,\dotso,x_K\right)\right.$ $\left.\bigm\vert f(\hat{x}_1^R, \hat{x}_2^R,\hat{x}_K^R)= f\left(x_1,x_2,\dotso,x_K\right)\right\rbrace $ has a diversity order two.
\begin{proof}

Recall that the decoder used at D given in \eqref{eqn_decoder} in Section II A, involves computation of the metrics $m_1$ and $m_2$ defined in \eqref{metric_m1} and \eqref{metric_m2}. Under the condition that $R$ transmitted the correct network coding symbol, a decoding error occurs at $D$ only when $m_1(x_1,x_2,\dotso,x_K)>m_1(x'_1,x'_2,\dotso,x'_K)$ or $m_1(x_1,x_2,\dotso,x_K)>\log(SNR)+m_2(x'_1,x'_2,\dotso,x'_K)$ for some $(x'_1,x_2,\dotso,x'_K) \neq (x_1,x_2,\dotso,x_K).$ Hence, the probability {\small$P\left \lbrace(\hat{x}_1^D, \hat{x}_2^D,\dotso,\hat{x}_K^D) \neq \left(x_1,x_2,\dotso,x_K\right)\right.$ $\left.\bigm\vert f(\hat{x}_1^R, \hat{x}_2^R,\dotso,\hat{x}_K^R)= f\left(x_1,x_2,\dotso,x_K\right)\right\rbrace$} can be upper bounded as in \eqref{pe_latest_ub}, which can be upper bounded using the union bound as in \eqref{p1_H_ub2} (eqns. \eqref{pe_latest_ub} and \eqref{p1_H_ub2} are given at the next page).
{\begin{figure*}
\scriptsize
\begin{align}
\nonumber
&P\left \lbrace(\hat{x}_1^D, \hat{x}_2^D,\dotso,\hat{x}_K^D) \neq \left(x_1,x_2,\dotso,x_K\right) \bigm\vert f(\hat{x}_1^R, \hat{x}_2^R,\dotso,\hat{x}_K^R)= f\left(x_1,x_2,\dotso,x_K\right)\right\rbrace\\
\nonumber
&=P \left \lbrace \left \lbrace m_1(x_1,x_2,\dotso,x_K) > m_1(x'_1,x'_2,\dotso,x'_K), (x'_1,x'_2,\dotso,x'_K) \neq (x_1,x_2,\dotso,x_K)\right \rbrace \right.
\\
\label{pe_latest_ub}
&\hspace{1 cm}\left.\cup \left \lbrace m_1(x_1,x_2,\dotso,x_K) > \log(SNR)
+m_2(x'_1,x'_2,\dotso,x'_K), (x'_1,x'_2,\dotso,x'_K) \neq (x_1,x_2,\dotso,x_K) \right \rbrace \bigm\vert f(\hat{x}_1^R, \hat{x}_2^R,\dotso,\hat{x}_K^R)= f\left(x_1,x_2,\dotso,x_K \right)\right \rbrace\\
\nonumber
& \hspace{-0 cm}\leq\hspace{-.7 cm}\sum_{\substack{{(x'_1,x'_2,\dotso,x'_K) \in \mathcal{S}^K}\\{(x'_1,x'_2,\dotso,x'_K)\neq (x_1,x_2,\dotso,x_K)}}} \hspace{-.7 cm}P \left \lbrace m_1(x_1,x_2,\dotso,x_K)> m_1(x'_1,x'_2,\dotso, x'_K)\bigm\vert f(\hat{x}_1^R, \hat{x}_2^R,\hat{x}_K^R)= f\left(x_1,x_2,\dotso,x_K\right)\right \rbrace \\
\label{p1_H_ub2}
&\hspace{2.1 cm}+ \hspace{-0.7 cm}\sum_{\substack{{(x'_1,x'_2,x'_K) \in \mathcal{S}^K}\\{(x'_1,x'_2,\dotso,x'_K)\neq (x_1,x_2,\dotso,x_K)}}}\hspace{-.7 cm} P\left \lbrace m_1(x_1,x_2,\dotso,x_K)> \log(SNR)+m_2(x'_1,x'_2,\dotso,x'_K)\bigm\vert f(\hat{x}_1^R, \hat{x}_2^R,\dotso,\hat{x}_K^R)= f\left(x_1,x_2,\dotso,x_K\right)\right \rbrace.
\end{align}
\hrule
\end{figure*}
}

 The probability {\small $P\left \lbrace m_1(x_1,x_2,\dotso,x_K)> m_1(x'_1,x'_2,\dotso,x'_K)\right.$ $\left. \bigm\vert f(\hat{x}_1^R, \hat{x}_2^R,\dotso,\hat{x}_K^R)= f\left(x_1,x_2,\dotso,x_K\right)\right \rbrace$} is equal to the Pair-wise Error Probability (PEP) of a space time coded $3 \times 1$ collocated MISO system, with the codeword difference matrices of the space time code used at the transmitter being of the form {\scriptsize $\begin{bmatrix} a_1 \Delta x_1 & a_2 \Delta x_2 &\dotso& a_K \Delta x_K & 0 \\ b_1 \Delta x_1 & b_2 \Delta x_2 &\dotso& b_K \Delta x_K  & \Delta x_R \end{bmatrix}^T,$} where $\Delta x_i=x_i-x'_i, i \in \lbrace 1,2,\dotso, K$ and $\Delta x_R=f(x_1,x_2,\dotso,x_K)-f(x'_1,x'_2,\dotso, x'_K).$ When $\Delta x_i \neq 0,$ for at least two values of $i \in \lbrace 1,2,\dotso, K \rbrace,$ these codeword difference matrices are of rank 2, other wise condition 1) given in the statement of Theorem 1) will be violated. When $\Delta x_i \neq 0$ and $\Delta x_j = 0, \forall i  \neq j, i, j \in \lbrace 1,2,\dotso, K \rbrace,$ the codeword difference matrices are full rank, otherwise condition 2) in the statement of Theorem 1 will be violated. Since the codeword difference matrices are full rank, the probability {\small$P\left \lbrace m_1(x_1,x_2,\dotso,x_K)> m_1(x'_1,x'_2,\dotso,x'_K)\right.$ $\left.\bigm\vert f(\hat{x}_1^R, \hat{x}_2^R,\dotso,\hat{x}_K^R)= f\left(x_1,x_2,\dotso,x_K\right)\right \rbrace$} has a diversity order two \cite{TaSeCa}.
 
\begin{figure*}
\scriptsize
\begin{align}
\label{metric_m4}
&m_4(x_1,x_2,\dotso,x_K,x_R)=\left\vert y_{D_1}-\sum_{i=1}^K h_{S_iD} \sqrt{E_s} a_i \: x_i\right\vert^2+ \left\vert y_{D_2} -\sum_{i=1}^K h_{S_iD} \sqrt{E_s} b_i \: x_i- h_{RD} \sqrt{E_s} x_R \right\vert ^2+\log(SNR).
\end{align}
\hrule
\end{figure*}
{
\begin{figure*}
\scriptsize
\begin{align}
\nonumber
&P \left \lbrace m_1(x_1,x_2,\dotso,x_K)> m_2(x'_1,x'_2,\dotso,x'_K)+\log(SNR) \bigm\vert f(\hat{x}_1^R, \hat{x}_2^R,\dotso,\hat{x}_K^R)= f\left(x_1,x_2,\dotso,x_K\right)\right \rbrace \\
\label{eqn_m1_m4}
&\hspace{4.5 cm}=P \left \lbrace m_1(x_1,x_2,\dotso,x_K)> \hspace{-.5 cm}\min_{x'_R \neq f(x'_1,x'_2,\dotso,x'_K)} m_4(x'_1,x'_2,\dotso,x'_K,x'_R) \bigm\vert f(\hat{x}_1^R, \hat{x}_2^R,\dotso,\hat{x}_K^R)= f\left(x_1,x_2,\dotso,x_K\right)\right \rbrace\\
\label{p1_ub4}
&\hspace{4.5 cm}\leq\hspace{-.5 cm} \sum_{x'_R \neq f(x'_1,x'_2,\dotso,x'_K)} \hspace{-.5 cm}P \left \lbrace m_1(x_1,x_2,\dotso,x_K)> m_4(x'_1,x'_2,\dotso,x'_K,x'_R) \bigm\vert f(\hat{x}_1^R, \hat{x}_2^R,\dotso,\hat{x}_K^R)= f\left(x_1,x_2,\dotso,x_K\right)\right \rbrace.\\
\hline
\nonumber
&\hspace{-0 cm}P_H \left \lbrace m_1(x_1,x_2,\dotso,x_K)>m_4(x'_1,x'_2,\dotso,x'_K,x'_R)\bigm\vert f(\hat{x}_1^R, \hat{x}_2^R,\dotso,\hat{x}_K^R) = f(x_1,x_2,\dotso,x_K)\right \rbrace\\
\label{pe_eqn_first}
&\hspace{-0cm}= P_H\left \lbrace \right \vert z_{D_1}\vert^2+\vert z_{D_2}\vert^2
 > \log(SNR)+\vert z_{D_1}+\sum_{i=1}^{K}h_{s_iD}\sqrt{E_s}a_i\Delta x_i \vert^2+\vert z_{D_2}+\sum_{i=1}^K h_{S_iD}\sqrt{E_s}b_i \Delta x_i +h_{RD}\sqrt{E_s}\Delta x_R\vert^2\rbrace.\\
\label{pe_eqn_1_first}
&\hspace{-0cm}=P_{H}\left\lbrace 2Re\left\lbrace \mathbf{z_D}^*\frac{\mathbf{\texttt{x}}}{\left \|\mathbf{\texttt{x}}\right \|}\right \rbrace \leq \frac{-\log\left(SNR\right)}{\left\|\mathbf{\texttt{x}}\right\|}-{\left\|\mathbf{\texttt{x}}\right\|}\right\rbrace=P_{H}\left\lbrace w \leq \frac{-\log\left(SNR\right)}{\sqrt{2}\left\|\mathbf{\texttt{x}}\right\|}-\frac{\left\|\mathbf{\texttt{x}}\right\|}{\sqrt{2}}\right\rbrace.
\end{align}
\hrule
\end{figure*}
} 

Let $m_4$ be a metric as defined in \eqref{metric_m4}, given in the next page. The probability {\small$P \left \lbrace m_1(x_1,x_2,\dotso,x_K)> m_2(x'_1,x'_2,\dotso,x'_K)+\log(SNR)\right.$ $\left.\bigm\vert\right.$ $\left.f(\hat{x}_1^R, \hat{x}_2^R,\dotso,\hat{x}_K^R)= f\left(x_1,x_2,\dotso,x_K\right)\right \rbrace$} can be written in terms of the metrics $m_1$ and $m_4$ as in \eqref{eqn_m1_m4}, which can be upper bounded as in \eqref{p1_ub4} (eqns. \eqref{eqn_m1_m4} --  \eqref{pe_eqn_1_first} are given in the next page).

Let $\Delta x_R=f(x_1,x_2,\dotso,x_K)-x'_R,$ $\Delta x_i=x_i-x'_i.$
The probability {\small $P_H \left \lbrace m_1(x_1,x_2,\dotso,x_K)>m_4(x'_1,x'_2,\dotso,x'_K,x'_R)\right.$ $\left.\bigm\vert f(\hat{x}_1^R, \hat{x}_2^R,\dotso,\hat{x}_K^R) = f(x_1,x_2,\dotso,x_K)\right \rbrace$} can be written in terms of the additive noise $z_{D_1}$ and $z_{D_2},$ as given in \eqref{pe_eqn_first}.

Let $x_1=\sum_{i=1}^K h_{S_iD}a_i\sqrt{E_s}\Delta x_i,$ $x_2=(\sum_{i=1}^K (h_{S_iD}b_i+h_{RD} \Delta x_R)\sqrt{E_s}.$ Also, let $\mathbf{z_D}=[z_{D_1}\; z_{D_2}]$ and $\mathbf{\texttt{x}}=[x_1 \; x_2]^T.$
Then \eqref{pe_eqn_first} can be simplified as in \eqref{pe_eqn_1_first}, where $w=\sqrt{2}Re\{\mathbf{z_D^*}\frac{\mathbf{\texttt{x}}}{\|\mathbf{\texttt{x}} \|}\},$ is distributed according to $\mathcal{N}(0,1).$  In terms of the $Q$ function, the probability $P_{H}\left\lbrace w \leq \frac{-\log\left(SNR\right)}{\sqrt{2}\left\|\mathbf{\texttt{x}}\right\|}-\frac{\left\|\mathbf{\texttt{x}}\right\|}{\sqrt{2}}\right\rbrace$ in \eqref{pe_eqn_1_first} can be written as $Q \left[ \frac{\log\left(SNR\right)}{\sqrt{2}\left\|\mathbf{\texttt{x}}\right\|}+\frac{\left\|\mathbf{\texttt{x}}\right\|}{\sqrt{2}}\right].$ Note that $\left\|\mathbf{\texttt{x}}\right\|$ depends on the fade coefficients. To complete the proof, it suffices to show that $\mathbb{E}\left(Q \left[ \frac{\log\left(SNR\right)}{\sqrt{2}\left\|\mathbf{\texttt{x}}\right\|}+\frac{\left\|\mathbf{\texttt{x}}\right\|}{\sqrt{2}}\right]\right)$ has a diversity order two.

The vector $\mathbf{\texttt{x}}$ can be written as,
\begin{align*}
\mathbf{\texttt{x}}&=\sqrt{E_s}\underbrace{[h_{S_1D}\; h_{S_2D}\;\dotso\;h_{S_KD} h_{RD}]}_{\mathbf{h}}\\
&\hspace{2.5 cm}\underbrace{ \begin{bmatrix} a_1 \Delta x_1 & a_2 \Delta x_2 &\dotso& a_K \Delta x_K & 0 \\ b_1 \Delta x_1 & b_2 \Delta x_2 &\dotso& b_K \Delta x_K  & \Delta x_R \end{bmatrix}^T}_{\mathbf{\Delta X}}.
\end{align*}

Since $\mathbf{\Delta X \Delta X^*}$ is Hermitian, it is unitarily diagonalizable, i.e, $ \mathbf{\Delta X \Delta X^*}=\mathbf{U}\mathbf{\Sigma}\mathbf{U^*},$ where $\mathbf{U}$ is unitary and $\mathbf{\Sigma}$ is a diagonal matrix. Since $\mathbf{\Delta X}$ has a maximum rank two, the number of non-zero diagonal entries of $\Sigma$ has to be less than or equal to two. Let $\lambda_1$ and $\lambda_2$ denote the two diagonal entries of $ \Sigma$ which are possibly non-zero, with $\lambda_1 \geq \lambda_2.$  We have $\left\|\mathbf{\texttt{x}}\right\|^2=SNR\;\mathbf{h}\mathbf{U}\mathbf{\Sigma}\mathbf{U}^*\mathbf{h}^*.$ Let $\mathbf{\tilde{h}}=\mathbf{h}\mathbf{U}=[\tilde{h}_1 \; \tilde{h}_2 \; \tilde{h}_3].$ The vector $\mathbf{\tilde{h}}$ has the same distribution as that of $\mathbf{h},$ since $\mathbf{U}$ is unitary.

Since the rank of $\mathbf{\Delta X}$ is at least one, $\lambda_1>0.$ We consider the two cases where $\lambda_2>0$ and $\lambda_2=0.$\\
\textit{\textbf{\noindent \hspace{-.5  cm}Case 1:}} $\lambda_2>0$\\
For this case, upper bounding $Q \left[ \frac{\log\left(SNR\right)}{\sqrt{2}\left\|\mathbf{\texttt{x}}\right\|}+\frac{\left\|\mathbf{\texttt{x}}\right\|}{\sqrt{2}}\right]$ by $Q \left[\frac{\left\|\mathbf{\texttt{x}}\right\|}{\sqrt{2}}\right],$ which is upper bounded by $e^{-\frac{\left\|\mathbf{\texttt{x}}\right\|^2}{4}},$ we have
\begin{align}
\label{eqn_eqn1}
\hspace{-.29 cm}Q \left[ \frac{\log\left(SNR\right)}{\sqrt{2}\left\|\mathbf{\texttt{x}}\right\|}+\frac{\left\|\mathbf{\texttt{x}}\right\|}{\sqrt{2}}\right] \leq e^{-\frac{1}{4}(\lambda_1 SNR \vert \tilde{h}_1 \vert^2+\lambda_2 SNR  \vert \tilde{h}_2 \vert^2)}.
\end{align}
Taking expectation w.r.t $\vert \tilde{h}_1 \vert$ and $\vert \tilde{h}_2 \vert,$ from \eqref{eqn_eqn1}, we get, 
  $\mathbb{E}\left(Q \left[ \frac{\log\left(SNR\right)}{\sqrt{2}\left\|\mathbf{\texttt{x}}\right\|}+\frac{\left\|\mathbf{\texttt{x}}\right\|}{\sqrt{2}}\right]\right) \leq \frac{1}{\left(1+\frac{\lambda_1 SNR}{4}\right)\left(1+\frac{\lambda_2 SNR}{4}\right)}.$ Hence $\mathbb{E}\left(Q \left[ \frac{\log\left(SNR\right)}{\sqrt{2}\left\|\mathbf{\texttt{x}}\right\|}+\frac{\left\|\mathbf{\texttt{x}}\right\|}{\sqrt{2}}\right]\right)$ has a diversity order two.
  
\textit{\textbf{\noindent \hspace{-.5  cm}Case 2:}} $\lambda_2=0$\\
 For this case $ \left\|\mathbf{\texttt{x}}\right\|=\sqrt{\lambda_1 SNR} \vert \tilde{h}_1\vert.$ Hence, 
 \begin{align}
 \nonumber
 Q \left[ \frac{\log\left(SNR\right)}{\sqrt{2}\left\|\mathbf{\texttt{x}}\right\|}+\frac{\left\|\mathbf{\texttt{x}}\right\|}{\sqrt{2}}\right]&\\
 \label{eqn_eqn2} 
 &\hspace{-2.5 cm}=Q \left[ \frac{\log\left(SNR\right)}{\sqrt{2}\sqrt{\lambda_1 SNR }\vert \tilde{h}_1\vert}+\frac{\sqrt{\lambda_1 SNR }\vert \tilde{h}_1\vert}{\sqrt{2}}\right].
 \end{align}

Let $r=\vert \tilde{h}_1\vert.$ Taking expectation w.r.t $r,$ from \eqref{eqn_eqn2}, we get,

{\small
\begin{align}
\nonumber
\mathbb{E}\left(Q \left[ \frac{\log\left(SNR\right)}{\sqrt{2}\left\|\mathbf{\texttt{x}}\right\|}+\frac{\left\|\mathbf{\texttt{x}}\right\|}{\sqrt{2}}\right]\right)&\\
\nonumber
&\hspace{-3.8 cm}=\int_{r=0}^{\infty} Q \left[ \frac{\log\left(SNR\right)}{\sqrt{2}\sqrt{\lambda_1 SNR }r}+\frac{\sqrt{\lambda_1 SNR }r}{\sqrt{2}}\right] 2r e^{-{r^2}} dr\\
\nonumber
&\hspace{-3.8 cm}=\underbrace{\int_{r=0}^{\sqrt{\frac{\log(SNR)}{\lambda_1 SNR}}} 2Q \left[ \frac{\log\left(SNR\right)}{\sqrt{2}\sqrt{\lambda_1 SNR }r}+\frac{\sqrt{\lambda_1 SNR }r}{\sqrt{2}}\right] r e^{-{r^2}} dr}_{I_1}\\
\nonumber
 &\hspace{-3.5 cm}+\underbrace{\int_{r=\sqrt{\frac{\log(SNR)}{\lambda_1 SNR}}}^{\infty} 2Q \left[ \frac{\log\left(SNR\right)}{\sqrt{2}\sqrt{\lambda_1 SNR }r}+\frac{\sqrt{\lambda_1 SNR }r}{\sqrt{2}}\right] r e^{-{r^2}} dr}_{I_2}.
\end{align}}

In the rest of the proof, we show that the integrals $I_1$ and $I_2$ have diversity order two. Note that $ \frac{\log\left(SNR\right)}{\sqrt{2}\sqrt{\lambda_1 SNR }r}+\frac{\sqrt{\lambda_1 SNR }r}{\sqrt{2}},$ as a function of $r,$ attains the minimum value when $r={\sqrt{\frac{\log(SNR)}{\lambda_1 SNR}}}$ and the minimum value equals $\sqrt{2\log(SNR)}.$ Since, $Q(x)$ is a decreasing function of $x,$ we have, $Q\left[\frac{\log\left(SNR\right)}{\sqrt{2}\sqrt{\lambda_1 SNR }r}+\frac{\sqrt{\lambda_1 SNR }r}{\sqrt{2}}\right]\leq Q \left[\sqrt{2\log(SNR)}\right].$ Hence, we have,
\begin{align*}
I_1 &\leq 2Q\left[\sqrt{2\log(SNR)}\right] \int_{r=0}^{\sqrt{\frac{\log(SNR)}{\lambda_1 SNR}}} r e^{-{r^2}} dr\\
&\leq \frac{2}{SNR}\left(1-e^{-\frac{\log(SNR)}{\lambda_1 SNR}}\right).
\end{align*}

Since for small $x,$ $e^{-x}$ can be approximated as $1-x,$ at high $SNR,$ we have 
$
I_1 \leq\frac{2}{SNR}\frac{\log(SNR)}{\lambda_1 SNR}
.$ Since $\displaystyle{\lim_{SNR \rightarrow \infty} \frac{-\log \left(\frac{2\log(SNR)}{\lambda_1 {SNR}^2}\right)}{\log(SNR)}=2},$ $I_1$ has a diversity order at least two.

Let $r_0=\sqrt{\frac{\log(SNR)}{\lambda_1 SNR}}.$
The integral $I_2$ can be upper bounded as,
$I_2\leq{\int_{r=r_0}^{\infty} Q \left[ \frac{\log\left(SNR\right)}{\sqrt{2}\sqrt{\lambda_1 SNR }r}+\frac{\sqrt{\lambda_1 SNR }r}{\sqrt{2}}\right] r  dr}.
$ Let $r'=\frac{\log\left(SNR\right)}{\sqrt{2}\sqrt{\lambda_1 SNR }r}+\frac{\sqrt{\lambda_1 SNR }r}{\sqrt{2}}.$ 
As a function of $r,$ $r'$ is monotonically increasing for $r \geq r_0.$ Also, for $r \geq r_0,$ $r$ can be written in terms of $r'$ as, $r=\frac{r'+\sqrt{{r'}^2-2\log(SNR)}}{\sqrt{2 \lambda_1 SNR}}.$ 
We have, $d {r}= d{r'} \frac{1}{\sqrt{2\lambda_1 SNR}} \left(1+\frac{r'}{\sqrt{{r'}^2-2\log(SNR)}}\right).$ Since $r \leq \frac{2r'}{\sqrt{2 \lambda_1 SNR}},$
$I_2$ can be upper bounded in terms of $r'$ as,

{\small
\begin{align*}
I_2&\leq{\int_{\sqrt{2\log(SNR)}}^{\infty}}  \frac{Q(r')r'}{{ \lambda_1 SNR}}  \left(1+\frac{r'}{\sqrt{{r'}^2-2\log(SNR)}}\right) d{r'}\\
&=\underbrace{\frac{1}{{\lambda_1 SNR}}\int_{\sqrt{2\log(SNR)}}^{\infty} Q(r') r' d{r'}}_{I_{21}}\\
&\hspace{1.6 cm}+\underbrace{\frac{1}{ \lambda_1 SNR}\int_{\sqrt{2\log(SNR)}}^{\infty} \frac{Q(r'){r'}^2}{\sqrt{{r'}^2-2\log(SNR)}} d {r'}}_{I_{22}}.
\end{align*}}

Upper bounding $Q(r')$ by $e^{-\frac{r'^2}{2}},$ $I_{21}$ can be shown to be upper bounded as $\frac{2}{\lambda_1 SNR^2},$ which falls as ${SNR}^{-2}.$ Upper bounding $Q(r')$ by $e^{-\frac{r'^2}{2}},$ and using the transformation $t={r'}^2-2\log(SNR),$ $I_{22}$ can be upper bounded as,
\begin{align*}
I_{22} &\leq \frac{1}{\lambda_1 SNR^2}\int_{0}^{\infty} \frac{t e^{-\frac{t}{2}}}{\sqrt{t}{\sqrt{t+2\log(SNR)}}} dt \\
&  \hspace{1.8 cm} + \frac{2\log(SNR)}{\lambda_1 SNR^2}\int_{0}^{\infty} \frac{ e^{-\frac{t}{2}}}{\sqrt{t}{\sqrt{t+2\log(SNR)}}} dt\\
& \leq \frac{1}{\lambda_1 SNR^2}\int_{0}^{\infty} e^{-\frac{t}{2}} dt + \frac{1}{\lambda_1 SNR^2}\int_{0}^{\infty} \frac{e^{-\frac{t}{2}}}{\sqrt{t}} dt\\
&=\frac{2+2\sqrt{2\pi}\log(SNR)}{\lambda_1 SNR^2}.
\end{align*}
where the second inequality above follows from the facts that $\frac{1}{\sqrt{t+2 \log(SNR)}} \leq \frac{1}{\sqrt{t}}$ and $\frac{1}{\sqrt{t+2 \log(SNR)}} \leq 1$ for sufficiently large $SNR.$ The last equality follows from the fact that $\int_{0}^{\infty} \frac{e^{-\frac{t}{2}}}{\sqrt{t}} dt=\sqrt{2}\Gamma(1/2)=\sqrt{2\pi},$ where $\Gamma(z)$ is the integral, $\Gamma(z)=\int_{0}^{\infty} e^{-t} t^{z-1} dt.$ Since, $\displaystyle\lim_{SNR \rightarrow \infty} \frac{-\log \left(\frac{2+2\sqrt{2\pi}\log(SNR)}{\lambda_1 SNR^2}\right)}{\log(SNR)}=2,$ $I_{22}$ has a diversity order 2. This completes the proof of Lemma 1.
\end{proof}
\end{lemma}

\begin{lemma}
When the two conditions in the statement of Theorem 1 are satisfied, the probability $P\left \lbrace(\hat{x}_1^D , \hat{x}_2^D,\dotso,\hat{x}_K^D) \neq \left(x_1,x_2,\dotso,x_K\right) \bigm\vert \right.$ $\left.f(\hat{x}_1^R, \hat{x}_2^R,\dotso,\hat{x}_K^R)= x'_R\right\rbrace, x'_R \neq f(x_1,x_2,\dotso,x_K),$ has a diversity order one.
\begin{proof}
\begin{figure*}
\scriptsize
\begin{align}
\label{metric_m4_repeat}
&m_4(x_1,x_2,\dotso,x_K,x_R)=\left\vert y_{D_1}-\sum_{i=1}^K h_{S_iD} \sqrt{E_s} a_i \: x_i\right\vert^2+ \left\vert y_{D_2} -\sum_{i=1}^K h_{S_iD} \sqrt{E_s} b_i x_i- h_{RD} \sqrt{E_s} x_R \right\vert ^2+\log(SNR).
\end{align}
\hrule
\end{figure*}
Let $m_4$ denote the metric as defined in \eqref{metric_m4_repeat}, given at the top of the next page.
Under the condition that $R$ transmitted the wrong network coded symbol $x'_R,$ a decoding error occurs at $D$ only when $m_4(x_1,x_2,\dotso,x_K,x'_R)>m_4(x''_1,x''_2,\dotso,x''_K,x''_R)$ or $m_4(x_1,x_2,\dotso,x_K,x'_R)>m_1(x''_1,x''_2,\dotso,x''_K),$ for some $(x''_1,x''_2,\dotso,x''_K) \neq (x_1,x_2,\dotso,x_K)$ and $x''_R \neq f(x''_1,x''_2,\dotso,x''_K).$ Hence, the probability $P\left \lbrace(\hat{x}_1^D , \hat{x}_2^D,\dotso,\hat{x}_K^D) \neq \left(x_1,x_2,\dotso,x_K\right)\right.$ $\left. \bigm\vert f(\hat{x}_1^R, \hat{x}_2^R,\dotso,\hat{x}_K^R) = x'_R\right\rbrace$ can be upper bounded as in \eqref{pe_latest_ub_lemma2} (eqns. \eqref{pe_latest_ub_lemma2} -- \eqref{pe_eqn_1} are shown at the top of the next page). Using the union bound, from \eqref{pe_latest_ub_lemma2}, we get \eqref{pe_latest_ub1_lemma2}.

{\begin{figure*}
\scriptsize
\begin{align}
\nonumber
&P\left \lbrace(\hat{x}_1^D , \hat{x}_2^D,\dotso,\hat{x}_K^D) \neq \left(x_1,x_2,\dotso,x_K\right) \bigm\vert f(\hat{x}_1^R, \hat{x}_2^R,\dotso,\hat{x}_K^R) = x'_R\right\rbrace
\\
\label{pe_latest_ub_lemma2}
&=\hspace{-.7 cm}\sum_{\substack{{(x''_A,x''_B) \in \mathcal{S}^2}\\{(x''_A,x''_B)\neq (x_1,x_2,\dotso,x_K)}}} \hspace{-1 cm} P \left \lbrace \left \lbrace m_4(x_1,x_2,\dotso,x_K,x'_R) > m_4(x''_A,x''_B,x''_R), x''_R \neq f(x''_A,x''_B)\right \rbrace \right.\left.
\hspace{0 cm}\cup \left \lbrace m_4(x_1,x_2,\dotso,x_K,x'_R) > 
m_1(x''_A,x''_B) \right \rbrace \bigm\vert f(\hat{x}_1^R, \hat{x}_2^R,\dotso,\hat{x}_K^R)= x'_R\right \rbrace\\
\nonumber
&\leq \hspace{-.4 cm}\sum_{\substack{{(x''_A,x''_B) \in \mathcal{S}^2}\\{(x''_A,x''_B)\neq (x_1,x_2,\dotso,x_K)}}} \hspace{-.0cm}\sum_{\substack{{x''_R \in \mathcal{S},}\\{ x''_R \neq f(x''_A,x''_B)}}}\hspace{-.3 cm} P \left \lbrace m_4(x_1,x_2,\dotso,x_K,x'_R)>m_4(x''_A,x''_B,x''_R)\right.\left.\bigm\vert f(\hat{x}_1^R, \hat{x}_2^R,\dotso,\hat{x}_K^R) = x'_R\right\rbrace\\
\label{pe_latest_ub1_lemma2}
&\hspace{6.4 cm}+\hspace{-.57 cm}\sum_{\substack{{(x''_A,x''_B) \in \mathcal{S}^2}\\{(x''_A,x''_B)\neq (x_1,x_2,\dotso,x_K)}}}\hspace{-.5 cm} P \hspace{-.1 cm}\left \lbrace m_4(x_1,x_2,\dotso,x_K,x'_R)>m_1(x''_A,x''_B)\right. \left.\bigm\vert f(\hat{x}_1^R, \hat{x}_2^R,\dotso,\hat{x}_K^R) = x'_R\right\rbrace.
\end{align}
\hrule
\end{figure*}}

Since the matrix $\begin{bmatrix} a_1 \Delta x_1 & a_2 \Delta x_2 &\dotso& a_K \Delta x_K & 0 \\ b_1 \Delta x_1 & b_2 \Delta x_2 &\dotso& b_K \Delta x_K  & \Delta x_R \end{bmatrix}^T$
has rank at least one for $(x_1,x_2,\dotso,x_K) \neq (x''_1,x''_2,\dotso,x''_K),$ where $\Delta x_R=x'_R-f(x''_1,x''_2,\dotso,x''_K),$ and $\Delta x_i= x_i-x''_i,$ the probability $P \left \lbrace m_4(x_1,x_2,\dotso,x_K,x'_R)>m_4(x''_1,x''_2,\dotso,x''_K,x''_R)\right.$ $\left.\bigm\vert f(\hat{x}_1^R, \hat{x}_2^R,\dotso,\hat{x}_K^R) = x'_R\right \rbrace$ has a diversity order at least one.
 
$P_H \left \lbrace m_4(x_1,x_2,\dotso,x_K,x'_R)>m_1(x''_1,x''_2,\dotso,x''_K)\right.$ $\left.\bigm\vert f(\hat{x}_1^R, \hat{x}_2^R,\dotso,\hat{x}_K^R) = x'_R\right \rbrace$ can be written in terms of the additive noise $z_{D_1}$ and $z_{D_2},$ as given in \eqref{pe_eqn}.

{\begin{figure*}
\scriptsize
\begin{align}
\nonumber
P_H \left \lbrace \log(SNR)+m_4(x_1,x_2,\dotso,x_K,x'_R)>m_1(x''_A,x''_B)\bigm\vert f(\hat{x}_1^R, \hat{x}_2^R,\dotso,\hat{x}_K^R) = x'_R\right \rbrace\\
\label{pe_eqn}
&\hspace{-8cm}= P_H\left \lbrace \right \vert z_{D_1}\vert^2+\vert z_{D_2}\vert^2+\log(SNR)
 > \vert z_{D_1}+\sum_{i=1}^K h_{S_iD}\sqrt{E_s}a_i\Delta x_i +\vert z_{D_2}+\sum_{i=1}^K h_{S_iD}\sqrt{E_s}b_i\Delta x_i +h_{RD}\Delta x_R\vert^2\rbrace.\\
\label{pe_eqn_1}
&\hspace{-8cm}=P_{H}\left\lbrace 2Re\left\lbrace \mathbf{z_D}^*\frac{\mathbf{\texttt{x}}}{\left \|\mathbf{\texttt{x}}\right \|}\right \rbrace \leq \frac{\log\left(SNR\right)}{\left\|\mathbf{\texttt{x}}\right\|}-{\left\|\mathbf{\texttt{x}}\right\|}\right\rbrace=P_{H}\left\lbrace w \leq \frac{\log\left(SNR\right)}{\sqrt{2}\left\|\mathbf{\texttt{x}}\right\|}-\frac{\left\|\mathbf{\texttt{x}}\right\|}{\sqrt{2}}\right\rbrace.
\end{align}
\hrule
\end{figure*}}

Let $x_1=\sum_{i=1}^{K}h_{S_iD}a_i\sqrt{E_s}x_i,$ $x_2=(\sum_{i=1}^K h_{S_iD}b_i x_i+h_{RD} \Delta x_R)\sqrt{E_s}.$ Also, let $\mathbf{z_D}=[z_{D_1}\; z_{D_2}]$ and $\mathbf{\texttt{x}}=[x_1 \; x_2]^T.$ Then \eqref{pe_eqn} can be simplified as in \eqref{pe_eqn_1},
 where $w=\sqrt{2}Re\{\mathbf{z_D^*}\frac{\mathbf{\texttt{x}}}{\|\mathbf{\texttt{x}} \|}\},$ is distributed according to $\mathcal{N}(0,1).$ Hence, 
{
\begin{align}
\nonumber
&P_{H}\hspace{-.1 cm}\left \lbrace  w \leq \frac{\log\left(SNR\right)}{\sqrt{2}\left\|\mathbf{\texttt{x}}\right\|}-\frac{{\left\|\mathbf{\texttt{x}}\right\|}}{\sqrt{2}}\right\rbrace\\
\nonumber
&= 1_{\{{\left\|\mathbf{\texttt{x}}\right\|^2}\leq \log\left(SNR\right)\}}\left(1-Q\left[\frac{\log\left(SNR\right)}{\sqrt{2}\left\|\mathbf{\texttt{x}}\right\|}-\frac{\left\|\mathbf{\texttt{x}}\right\|}{\sqrt{2}}\right]\right)\\
\label{pe_eqn2}
&\hspace{1.05 cm}+1_{\{{\left\|\mathbf{\texttt{x}}\right\|^2}> \log\left(SNR\right)\}}Q\left[-\frac{\log\left(SNR\right)}{\sqrt{2}\left\|\mathbf{\texttt{x}}\right\|}+\frac{\left\|\mathbf{\texttt{x}}\right\|}{\sqrt{2}}\right].
\end{align}}
Taking expectation with respect to the fade coefficients in \eqref{pe_eqn2}, $P\left \lbrace w \leq \frac{\log\left(SNR\right)}{\sqrt{2}\left\|\mathbf{\texttt{x}}\right\|}-\frac{\left\|\mathbf{\texttt{x}}\right\|}{\sqrt{2}}\right\rbrace$ can be upper bounded as,
{
\begin{align}
\nonumber
&\hspace{-0 cm}P\left \lbrace w \leq \frac{\log\left(SNR\right)}{\sqrt{2}\left\|\mathbf{\texttt{x}}\right\|}-\frac{\left\|\mathbf{\texttt{x}}\right\|}{\sqrt{2}}\right\rbrace \leq P\left \lbrace {\left\|\mathbf{\texttt{x}}\right\|}^2 \leq \log(SNR) \right \rbrace\\
\label{pe_eqn3}
&\hspace{0.2 cm}+\int_{H:\{{\left\|\mathbf{\texttt{x}}\right\|^2}> \log\left(SNR\right)\}} Q\left[-\frac{\log\left(SNR\right)}{\sqrt{2}\left\|\mathbf{\texttt{x}}\right\|}+\frac{\left\|\mathbf{\texttt{x}}\right\|}{\sqrt{2}}\right] dH.
\end{align}}
The vector $\mathbf{\texttt{x}}$ can be written as,
\begin{align*}
\mathbf{\texttt{x}}&=\sqrt{E_s}\underbrace{[h_{S_1D}\; h_{S_2D}\;\dotso\;h_{S_KD} h_{RD}]}_{\mathbf{h}}\\
&\hspace{2.5 cm}\underbrace{ \begin{bmatrix} a_1 \Delta x_1 & a_2 \Delta x_2 &\dotso& a_K \Delta x_K & 0 \\ b_1 \Delta x_1 & b_2 \Delta x_2 &\dotso& b_K \Delta x_K  & \Delta x_R \end{bmatrix}^T}_{\mathbf{\Delta X}}.
\end{align*} 
Since $\mathbf{\Delta X \Delta X^*}$ is Hermitian, it is unitarily diagonalizable, i.e, $ \mathbf{\Delta X \Delta X^*}=\mathbf{U}\mathbf{\Sigma}\mathbf{U^*},$ where $\mathbf{U}$ is unitary and $\mathbf{\Sigma}$ is diagonal with $\lambda_1$ and $\lambda_2$ denoting the two possible non-zero diagonal entries, where $\lambda_1 \geq \lambda_2.$ Since the rank of $\mathbf{\Delta X}$ is at least one, $\lambda_1>0.$ We have $\left\|\mathbf{\texttt{x}}\right\|^2=SNR\;\mathbf{h}\mathbf{U}\mathbf{\Sigma}\mathbf{U}^*\mathbf{h}^*.$ Let $\mathbf{\tilde{h}}=\mathbf{h}\mathbf{U}=[\tilde{h}_1 \; \tilde{h}_2 \; \tilde{h}_3].$ The vector $\mathbf{\tilde{h}}$ has the same distribution as that of $\mathbf{h},$ since $\mathbf{U}$ is unitary. Hence, we have,
{\small
\begin{align}
\nonumber 
P\left \lbrace {\left\|\mathbf{\texttt{x}}\right\|}^2 \leq \log(SNR) \right \rbrace&=P\left \lbrace \lambda_1 \vert \tilde{h}_1\vert^2+\lambda_2 \vert \tilde{h}_2\vert^2 \leq \frac{\log(SNR)}{SNR}\right \rbrace\\
\nonumber
& \leq P\left \lbrace \lambda_1 \vert \tilde{h}_1\vert^2 \leq \frac{\log(SNR)}{SNR}\right \rbrace.
\end{align}
}
Since, $\vert \tilde{h}_1 \vert ^2$ is exponentially distributed, 
{
\begin{align}
\nonumber 
P\left \lbrace {\left\|\mathbf{\texttt{x}}\right\|}^2 \leq \log(SNR) \right \rbrace \leq \left(1-e^{-\frac{\log(SNR)}{\lambda_1 SNR}}\right).
\end{align}}
At high SNR, $1-e^{-\frac{\log(SNR)}{\lambda_1 SNR}}$ can be approximated as $\frac{\log(SNR)}{\lambda_1 SNR}.$  $P\left \lbrace {\left\|\mathbf{\texttt{x}}\right\|}^2 \leq \log(SNR) \right \rbrace$ has a diversity order at least one since  $\displaystyle{\lim_{SNR \rightarrow \infty} \frac{-\log\left(\frac{\log(SNR)}{\lambda_1 SNR}\right)}{\log \left(SNR \right)}=1.}$ 
Since $Q(x)<e^{-\frac{x^2}{2}},$ the integral on the right hand side of \eqref{pe_eqn3} can be upper bounded as,
{
\begin{align}
\nonumber
&\int_{H:\{{\left\|\mathbf{\texttt{x}}\right\|^2}> \log\left(SNR\right)\}} Q\left[-\frac{\log\left(SNR\right)}{\sqrt{2}\left\|\mathbf{\texttt{x}}\right\|}+\frac{\left\|\mathbf{\texttt{x}}\right\|}{\sqrt{2}}\right] dH \\
\label{eqn_temp}
&\leq \int_{H:\{{\left\|\mathbf{\texttt{x}}\right\|^2}> \log\left(SNR\right)\}} e^{-\frac{\left(-\frac{\log\left(SNR\right)}{\left\|\mathbf{\texttt{x}}\right\|}+{\left\|\mathbf{\texttt{x}}\right\|}\right)^2}{4}} dH.
\end{align}
} 
We consider the following two cases when $\lambda_2>0$ and $\lambda_2=0.$\\
\textit{\textbf{\hspace{-.7 cm}Case 1:}} $\lambda_2>0$\\
For this case, from the integral in \eqref{eqn_temp}, we get,
{\small
\vspace{-.1 cm}
\begin{align}
\nonumber
&\int_{H:\{{\left\|\mathbf{\texttt{x}}\right\|^2}> \log\left(SNR\right)\}} Q\left[-\frac{\log\left(SNR\right)}{\sqrt{2}\left\|\mathbf{\texttt{x}}\right\|}+\frac{\left\|\mathbf{\texttt{x}}\right\|}{\sqrt{2}}\right] dH \\
\nonumber
&\leq \int_{\vert\tilde{h}_1\vert^2=0}^{\infty} \int_{\vert\tilde{h}_2\vert^2=0}^{\infty} e^{\log\left(SNR\right)} e^{-\frac{SNR(\lambda_1 \vert\tilde{h}_1\vert^2+\lambda_2 \vert\tilde{h}_2\vert^2)}{4}} \\
\nonumber
&\hspace{4.5 cm}e^{-\vert\tilde{h}_1\vert^2} e^{-\vert\tilde{h}_2\vert^2} d\vert\tilde{h}_1\vert^2\;d\vert\tilde{h}_2\vert^2\\
\nonumber
&=\frac{SNR}{\left({1+\frac{\lambda_1 SNR}{4}}\right)\left({1+\frac{\lambda_2 SNR}{4}}\right)}, 
\end{align}}which falls as $SNR^{-1}$ at high SNR.\\
\textit{\textbf{Case 2:}} $\lambda_2=0$\\
For this case,
{
\begin{align}
\nonumber
I \triangleq &\int_{H:\{{\left\|\mathbf{\texttt{x}}\right\|^2}> \log\left(SNR\right)\}} Q\left[-\frac{\log\left(SNR\right)}{\sqrt{2}\left\|\mathbf{\texttt{x}}\right\|}+\frac{\left\|\mathbf{\texttt{x}}\right\|}{\sqrt{2}}\right] f(H)dH \\
\nonumber
&\leq \int_{\vert\tilde{h}_1\vert >\frac{\sqrt{\log(SNR)}}{\sqrt{\lambda_1 \; SNR}}} Q\left[\frac{\sqrt{\lambda_1 SNR} \vert\tilde{h}_1\vert-\sqrt{\log(SNR)}}{\sqrt{2}}\right] \\
\label{eqn_temp2}
&\hspace{4.5 cm}2\vert \tilde{h}_1 \vert e^{-{\vert \tilde{h}_1 \vert^2}} d \vert\tilde{h}_1\vert.
\end{align}}
The above inequality follows from the fact that for $\vert\tilde{h}_1\vert >\frac{\sqrt{\log(SNR)}}{\sqrt{\lambda_1 \; SNR}},$ $-\frac{\log(SNR)}{\sqrt{\lambda_1 SNR}\vert \tilde{h}_1\vert}<-\sqrt{\log(SNR)}.$
Let $r=\vert \tilde{h}_1 \vert.$ From \eqref{eqn_temp2}, since $Q(x)<e^{-\frac{x^2}{2}}$ the integral $I$ can be upper bounded as,

{\vspace{-.7 cm}
\scriptsize
\begin{align}
\nonumber
&I \leq \int_{r >\underbrace{\frac{\sqrt{\log(SNR)}}{\sqrt{\lambda_1 \; SNR}}}_{r_0}} e^{-\underbrace{\left({1}+\frac{\lambda_1 SNR}{4}\right)}_{k_1}\left(r-\underbrace{\frac{\sqrt{\lambda_1 SNR \log(SNR)}}{4+\lambda_1 SNR}}_{k_2}\right)^2}\\
\label{eqn_temp3}
&\hspace{5 cm }\underbrace{e^{-\frac{\log(SNR)}{4}+\frac{\log(SNR)\lambda_1 SNR}{16+4\lambda_1 SNR}}}_{k_3}dr
\end{align}
\vspace{-0 cm}}

Let $r_0,$ $k_1,$ $k_2$ and $k_3$ be defined as shown in \eqref{eqn_temp3}. From \eqref{eqn_temp3}, the upper bound on $I$ can be written as,

{\small
\begin{align}
\nonumber
I \leq \frac{k_3}{2} \int_{r > r_0} 2(r-k_2) e^{-k_1(r-k_2)^2} dr + k_2\: k_3 \int_{r \geq r_0} e^{-k_1(r-k_2)^2} dr.
\end{align}
}

Since $\int_{r \geq r_0} e^{-k_1(r-k_2)^2} dr=\sqrt{\frac{\pi}{K}} Q \left[ (r_0-k_2)\sqrt{2k_1}\right],$ 
{
\begin{align}
\nonumber
I \leq \frac{k_3}{2 k_1} e^{-k_1(r_0-k_2)^2}+k_2\: k_3 \sqrt{\frac{\pi}{k_1}}Q \left[(r_0-k_2)\sqrt{2 k_1}\right].
\end{align}
}

 Since $Q \left[(r_0-k_2)\sqrt{2 k_1}\right] \leq 1,$ $e^{-k_1(r_0-k_2)^2}\leq 1,$ and $k_3$ can be approximated as one at high $SNR,$ substituting for $r_0,k_1,k_2$ and $k_3,$ we get,
{
\begin{align}
\nonumber
I \leq \frac{1}{2\left(1+\frac{\lambda_1 SNR}{4}\right)}+\frac{\sqrt{\pi\lambda_1 SNR \log(SNR)}}{\left(4+\lambda_1 SNR \right)\sqrt{{1}+\frac{\lambda_1 SNR}{4}}}.
\end{align}}

Since the above upper bound on $I$ falls as $SNR^{-1}$ at high SNR, $I$ has a diversity order at least 1. This completes the proof of Lemma 2.
\end{proof}
\end{lemma} 
\end{document}